%% file: main.tex
\documentclass[conference]{IEEEtran}
\usepackage{amsmath,amssymb,amsfonts}

\input{config.tex}
\IEEEoverridecommandlockouts
\usepackage{cite}
\usepackage{algorithmic}
\usepackage{graphicx}
\usepackage{textcomp}
\def\BibTeX{{\rm B\kern-.05em{\sc i\kern-.025em b}\kern-.08em
    T\kern-.1667em\lower.7ex\hbox{E}\kern-.125emX}}
\titlespacing{\subsubsection}
    {0pt}{5pt}{2pt} %left, above, below
\begin{document}
\pdfinclusioncopyfonts=1

\title{Adaptive Recursive Query Optimization}

\author{
    \IEEEauthorblockN{Anna Herlihy}
    \IEEEauthorblockA{
        \textit{EPFL}\\
        Switzerland \\
        anna.herlihy@epfl.ch
    }
    \and
    \IEEEauthorblockN{Guillaume Martres}
    \IEEEauthorblockA{
        \textit{EPFL}\\
        Switzerland \\
        guillaume.martres@epfl.ch
    }
    \and
    \IEEEauthorblockN{Anastasia Ailamaki}%\IEEEauthorrefmark{1}}
    \IEEEauthorblockA{%\textit{DIAS Lab} \\
        \textit{EPFL}\\
        Switzerland \\
        anastasia.ailamaki@epfl.ch
    }
    \and
    \IEEEauthorblockN{Martin Odersky}
    \IEEEauthorblockA{
        \textit{EPFL}\\
        Switzerland \\
        martin.odersky@epfl.ch
    }
}

\maketitle
\thispagestyle{plain}
\pagestyle{plain}
% \begingroup\renewcommand\thefootnote{\IEEEauthorrefmark{1}}
% \footnotetext{Work done in its entirety at EPFL.}
% \endgroup

\subimport*{00-abstract/}{main-v0.tex}

% \begin{IEEEkeywords}
% component, formatting, style, styling, insert
% \end{IEEEkeywords}

\setcounter{taskcnt}{0}\section{Introduction}\label{sec:intro}\subimport*{01-introduction/}{main-v3.tex}
\setcounter{taskcnt}{0}\section{Background}\label{sec:dl}\subimport*{02-background/}{main-v3.tex}
\setcounter{taskcnt}{0}\section{\arch{} Architecture}\label{sec:arch}\subimport*{03-arch/}{main-v0.tex}

\setcounter{taskcnt}{0}\section{Runtime Join-Order Optimization}\label{sec:opt}\subimport*{04-optimizations/}{main-v5.tex}
\setcounter{taskcnt}{0}\section{\system{}: System Design}\label{sec:sys}\subimport*{05-carac/}{main-v3.tex}
\setcounter{taskcnt}{0}\section{Evaluation}\label{sec:eval}\subimport*{06-eval/}{main-v3.tex}
\setcounter{taskcnt}{0}\section{Related work}\label{sec:rel}\subimport*{07-related-work/}{main-v0.tex}
\setcounter{taskcnt}{0}\section{Conclusion}\label{sec:conclusion}\subimport*{08-conclusion/}{main-v0.tex}

\section*{Acknowledgments}
We thank the anonymous reviewers, Amir Shaikhha, and the DIAS and LAMP labs for their insightful feedback. This work was supported in part by the SNSF Grant TMAG-2\_209506/1 ``Capabilities for Typing Resources and Effects".

% The preferred spelling of the word ``acknowledgment'' in America is without 
% an ``e'' after the ``g''. Avoid the stilted expression ``one of us (R. B. 
% G.) thanks $\ldots$''. Instead, try ``R. B. G. thanks$\ldots$''. Put sponsor 
% acknowledgments in the unnumbered footnote on the first page.

\bibliographystyle{IEEEtran}
\bibliography{bibliography}

\end{document}

%% file: config.tex
\providecommand{\subparagraph}{}

\usepackage[dvipsnames]{xcolor}
\usepackage{pgfplotstable}
\usepackage{lipsum}
\usepackage{balance}
\usepackage[defaultlines=3,all]{nowidow}
\usepackage{import}
\usepackage[inline]{enumitem}
\usepackage{titlesec}
\usepackage{chapterfolder}
\usepackage{graphicx}

\usepackage[hidelinks]{hyperref}

\usepackage{cleveref}
\usepackage{multirow}

\usepackage{fancyvrb}
\usepackage{siunitx}
\usepackage{tabularx}
\usepackage{makecell}

\noclub

\pgfplotsset{compat=1.9}
\usepgfplotslibrary{units}
\usepgfplotslibrary{colorbrewer}
\usetikzlibrary{patterns}

\usepackage{ifdraft}

\newcommand{\system}{Carac}
\newcommand{\arch}{Adaptive Metaprogramming}

\newcommand{\paragraphT}[1]{}
\newcommand{\punchlineT}[1]{}

\ifdraft{\let\ifxdraft\iftrue}{\let\ifxdraft\iffalse}

\definecolor[named]{ACMDarkBlue}{cmyk}{1,0.58,0,0.21}

\newcommand{\revA}[1]{{#1}}

\newcommand{\revB}[1]{{#1}}
\newcommand{\revC}[1]{{#1}}

\newcommand{\toberemoved}[1]{\ifdraft{}{}}

\titlespacing{\paragraph}{%
  0pt}{%              left margin
  0pt}{% space before (vertical)
  0.25em}%               space after (horizontal)

\titleclass{\subsubparagraph}{straight}[\subparagraph]
\newcounter{subsubparagraph}

\titleformat{\subsubparagraph}[runin]
  {\normalfont\normalsize}
  {\thesubsubparagraph.}
  {.5em}
  {}
\renewcommand\thesubsubparagraph{\roman{subsubparagraph}}
\titlespacing{\subsubparagraph}{%
  0pt}{%              left margin
  0pt}{% space before (vertical)
  0.25em}%               space after (horizontal)

\let\oldparagraph\paragraph
\newcommand{\paragraphnodot}[1]{\oldparagraph{\textbf{#1}}}
\renewcommand\paragraph[1]{\paragraphnodot{#1.}}

\let\oldsubparagraph\subparagraph
\newcommand{\subparagraphnodot}[1]{\oldsubparagraph{\emph{\underline{#1}}}}
\renewcommand\subparagraph[1]{\subparagraphnodot{#1.}}

\let\oldsubsubparagraph\subsubparagraph
\newcommand{\subsubparagraphnodot}[1]{\oldsubsubparagraph{\emph{#1}}}
\renewcommand\subsubparagraph[1]{\subsubparagraphnodot{\textbf{\S\S} #1.}}

\makeatletter
  \def\relativepath{\import@path}
\makeatother

\newcommand\notecolor{SeaGreen}

\newcounter{taskcnt}

\newcommand{\cfscratch}[4][]{%
\ifthenelse{\equal{#2}{}}{}{\ifthenelse{\equal{#1}{}}{{\color{\notecolor}\subsubsection*{#2}}}{{\color{\notecolor}\subsubsection*[#1]{#2}}}}%
{\color{\notecolor}\subimport*{#3/}{#4}}%
}
\newcommand{\cfparagraph}[4][]{%
\ifthenelse{\equal{#2}{}}{}{\ifthenelse{\equal{#1}{}}{\paragraph{#2}}{\paragraph[#1]{#2}}}%
\subimport*{#3/}{#4}%
}

\newcommand{\cfsubsubsection}[4][]{%
\setcounter{taskcnt}{0}%
\ifthenelse{\equal{#1}{}}{\subsubsection{#2}}{\subsubsection[#1]{#2}}%
\subimport*{#3/}{#4}%
}

\renewcommand{\cfparagraph}[4][]{%
\ifthenelse{\equal{#2}{}}{}{%
\ifthenelse{\equal{#1}{}}{\paragraph{#2}}{\paragraph[#1]{#2}}}%
\subimport*{#3/}{#4}%
}

\renewcommand{\cfsubsection}[4][]{%
\setcounter{taskcnt}{0}%
\ifthenelse{\equal{#1}{}}{\subsection{#2}}{\subsection[#1]{#2}}%
\subimport*{#3/}{#4}%
}

\renewcommand{\cfsection}[4][]{%
\setcounter{taskcnt}{0}%
\ifthenelse{\equal{#1}{}}{\section{#2}}{\section{#2}\label{#1}}%
\subimport*{#3/}{#4}%
}

\let\oldincludegraphics\includegraphics
\renewcommand{\includegraphics}[2][]{%
\oldincludegraphics[draft=false,#1]{\relativepath#2}%
}

%% file: 00-abstract/main-v0.tex
%!TEX root = ../main.tex
\begin{abstract}

Performance-critical industrial applications, including large-scale program, network, and distributed system analyses, are increasingly reliant on recursive queries for data analysis. Yet traditional relational algebra-based query optimization techniques do not scale well to recursive query processing due to the iterative nature of query evaluation, where relation cardinalities can change unpredictably during the course of a single query execution. To avoid error-prone cardinality estimation, adaptive query processing techniques use runtime information to inform query optimization, but these systems are not optimized for the specific needs of recursive query processing. % and (2) limited in the flexibility of runtime code generation. % due to the compilation mechanisms employed.
In this paper, we introduce \emph{\arch{}}, an innovative technique that 
% integrates Multi-Stage Programming with JIT data management systems, shifting 
shifts recursive query optimization and code generation from compile-time to runtime using \emph{principled metaprogramming}, enabling dynamic optimization and re-optimization before and after query execution has begun.
We present a custom join-ordering optimization 
% for conjunctive subqueries within recursive queries
applicable at multiple stages during query compilation and execution. 
% We introduce speculative execution and feedback loops into query execution, allowing for more scalable and adaptive query evaluation. 
Through \system{}, \revA{a custom Datalog engine}, we evaluate the optimization potential of \arch{}
% a system that not only tightly integrates declarative Datalog programs within a general-purpose programming language but also maintains performance competitiveness with advanced Datalog systems. We 
and show unoptimized recursive query execution time can be improved by three orders of magnitude and hand-optimized queries by 6x.

\end{abstract}

%% file: 01-introduction/main-v3.tex
% P: recursive queries gaining popularity
The advanced algorithms powering contemporary applications require increasingly expressive query capabilities beyond traditional relational algebra, for example iterative or fixed-point computations.
This demand has led to a resurgence in the use of recursion in analytical queries, yet the development of recursive query optimization techniques lags behind those established for non-recursive queries.

%P: why SQL sota is not enough
Support for recursion was added to the ANSI'99 SQL standard, and yet most relational data management systems are not optimized for the specific needs of recursive queries: during the execution of a recursive query, resulting relations from the current iteration are fed as input into the next iteration of query evaluation. As a result, the relation cardinalities can change unpredictably and the number of iterations required to execute the query is unknown ahead of time. Existing cardinality estimation techniques assume input relation cardinalities do not change during execution, \revA{and} so cannot guarantee high-quality join orders after the initial iteration.%\cite{souffle2}.

%P: instead, datalog, and why datalog sota isn't enough
Instead, developers turn to Datalog for high-performance recursive query processing. Datalog is a popular query language for recursive queries that 
has been used for Java program analysis~\cite{doop}, TensorFlow~\cite{tensorflow}, Ethereum Virtual Machine~\cite{etherium}, AWS network analysis~\cite{aws}, and other performance-critical applications. Most Datalog execution engines follow a bottom-up evaluation~\cite{ullman} strategy, in which subqueries in the form of unions of conjunctive queries are repeatedly applied to relations. 
State-of-the-art Datalog execution engines  
provide manual or \emph{offline} join-order optimizations for these subqueries, but require a profiling stage and cannot adapt if the real-time data does not match what was profiled.

%P: to avoid cardinality estimation, use AQP / codegen
To avoid error-prone cardinality estimation\cite{costmodels}, adaptive query processing (AQP) uses runtime information to tune query optimization \cite{aqp}. This can be combined with code-generating data management systems~\cite{vectorized,adaptiveexec,vida} that improve query execution performance by avoiding interpretation overhead by partially evaluating query plans into compiled executables. However, code-generating AQP systems 
do not regenerate and compile code at runtime, instead jumping between precompiled plans or code blocks to avoid the prohibitively high cost of invoking an optimizing compiler during query execution\cite{pcq,dblocks}. Because recursive queries require repeated execution of subqueries where results from the previous iteration are fed into the next iteration, an optimal query plan will not stay optimal over the execution of a single program and the number of potential paths is prohibitively large to enumerate or compile ahead-of-time. Thus, there is a need for query optimization techniques that can leverage runtime information to meet the specific optimization criteria of recursive query processing.
% and the limitations of the compilation mechanisms used to generate LLVM-IR.

To address these challenges, this work introduces \emph{\arch{}}, a novel technique for runtime optimization and repeated re-optimization of recursive queries. We implement and evaluate the design in \system{}, a system that tightly integrates a specialized Datalog execution engine with a general-purpose programming language and leverages Multi-Stage Programming\cite{gentlemsp}, a known technique from the Programming Languages community, to support runtime code generation. \revC{Ultimately, the goal of \system{} is to illustrate a novel optimization approach that leverages advanced metaprogramming language capabilities that would enable any recursive query processing engine to continuously reapply optimizations.} This paper makes the following contributions:
\begin{itemize}[align=parleft, topsep=0pt,leftmargin=*]

\item We introduce \arch{}, a technique that combines the power of code-generating just-in-time data management systems with Multi-Stage Programming to enable adaptive, continuous query re-optimization by regenerating code at runtime.
% push the optimization space for recursive queries from compile-time to runtime. 
%\arch{} expands query evaluation with speculation and feedback loops to allow Datalog execution to better scale and adapt to runtime information.
\item We leverage \arch{} to enable a custom 
optimization
% or cost mode?
for Datalog queries
% that combines selectivity estimation and relation cardinalities 
to select efficient join orders of conjunctive subqueries at different stages in the lifetime of the query: from ahead-of-time, when only queries are available, to just-in-time, where runtime statistics like relation cardinalities are available.
\item Through \revA{our custom Datalog engine} \system{}, we show how \arch{} can speed up unoptimized Datalog queries %1100x using only compile-time optimizations. Using only runtime optimizations, \system{} can speed up unoptimized queries by 
over 5000x while maintaining a 6x speedup over manually hand-optimized queries.

% \item \system{} provides a tight embedding of declarative Datalog programs within a general-purpose programming language, while maintaining competitive performance with state-of-the-art Datalog systems.
\end{itemize}

%% file: 02-background/main-v3.tex
\begin{figure}
    % \frame{
    \includegraphics[trim={1cm 1cm 1cm 1cm}, width=\linewidth,clip]{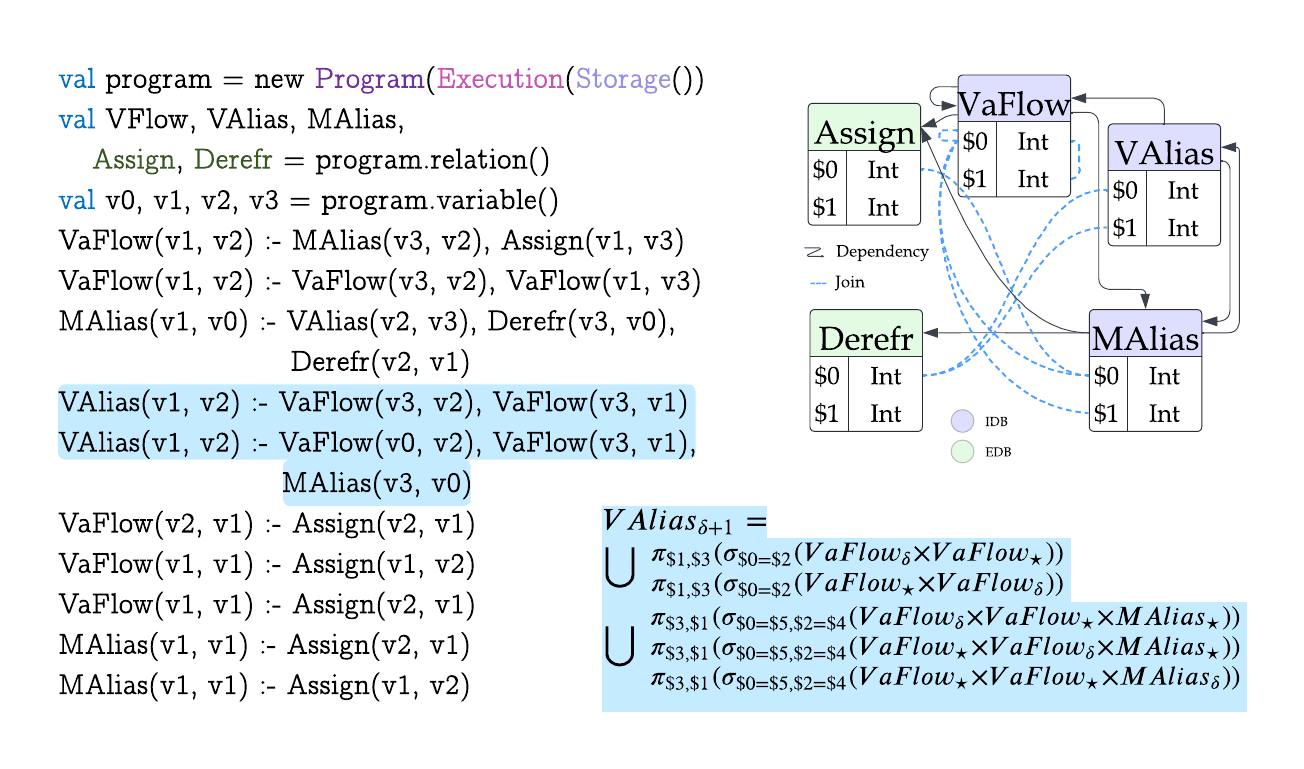}
    % }
    \caption{Graspan's Context-Sensitive Pointer Analysis (CSPA). Clockwise from left: (a) CSPA as a Carac Program. (b) EDB and IDB Database Schema illustrated with dependencies. (c) Example unnamed relational algebra query for the \texttt{VAlias} relation.}\label{fig:tc}
    % \vspace{-0.5cm}
\end{figure}
% \begin{figure}
%     \frame{
%     \includegraphics[trim={.1cm .1cm .1cm .1cm}, width=.6\linewidth,clip]{../figures/cspa-schema.pdf}
%     }
%     \caption{CSPA Schema}\label{fig:schema}
%     % \vspace{-0.5cm}
% \end{figure}

\subsection{Datalog}\label{msec:dl}
Datalog ~\cite{dl} is a rule-based declarative query language based in logic programming. Datalog programs take the form of a finite set of \textit{facts} and \textit{rules}. \revA{
% where rules define how new facts can be derived from existing facts. 
Facts are stored in the \emph{extensional database} (EDB) and rules are stored in the \emph{intensional database} (IDB).
% The rule schema is the relational schema of the intensional database, and the fact schema is the relational schema of the extensional database. 
A datalog program, or query, maps database instances over the fact schema (the relational schema of the EDB) to database instances over the rule schema (the relational schema of the IDB).} The bottom-up evaluation of a Datalog program iteratively computes subqueries containing unions of conjunctive queries for each derived predicate symbol until an iteration passes where no new facts are discovered\cite{dl}. 

\revA{
% For a Datalog rule of the form:
% % $R_0(v_0, ..., v_k)$:-$R_1(t_0,...,t_{k_1}), R_2(t_{k_1+1}, ..., t_{k_2}), ..., R_n(t_{k_{n-1}+1},...,t_{k_n})$
% where $v_i$ are variables, the subquery for $R$ run at each iteration takes the form of $\pi _V(\sigma _F(R_1 \Join _{C_1} ... \Join _{C_n} R_n))$. Columns are referred to by their ``position" in the body of the rule, i.e. $\$l$ is the position of term $t_l$ where $l \in [0..k_n]$. For each variable $v_i$ in the head, a position of $v_i$ in the body is added to $V$. For each $t_l$ that equals a constant $c$ the condition $\$l=c$ is added to $F$. If terms $t_l$ of relation $R_a$ and $t_m$ of relation $R_b$ equal the same variable, then the condition $R_a \Join _{\$l=\$m} R_b$ is added.
% }
For a datalog rule of the form:\\
$R_0(v_0, ..., v_k) \texttt{:-} R_1(t_0,...,t_{k_1}), ..., R_n(t_{k_{n-1}+1},...,t_{k_n})$
% $R_0(v_0, ..., v_k) \mathrel{\texttt{:-}} R_1(t_0,...,t_{k_1}), ..., R_n(t_{k_{n-1}+1},...,t_{k_n})$\\
where $v_k$ are variables, the generated query expressed in the unnamed relational algebra takes the form of $\pi _V(\sigma _F(R_1 \times ... \times R_n))$. $V$ contains the positions ($\$i$) of each $v_k$ in the body, and $F$ is a conjunction of conditions such that for each position $l$ in the body, if $t_l$ is a constant $c$, $F$ contains $\$l=c$. If $t_l$ is a variable that is present in a different position $m$, then $F$ contains $\$l=\$m$.  If there are multiple definitions for $R_0$, the result will be the union of the above expression for each rule definition. 
}
Evaluation can be made more efficient using the Semi-Naive evaluation technique, where only facts from the previous iteration are used to discover new facts~\cite{ullman}. This is done by splitting each query into $n$ unique subqueries where one of the $n$ relations is read from the ``delta" database, which contains only the facts discovered in the previous iteration, while the rest of the $n-1$ relations are read from the ``derived" database, which stores all facts discovered so far.

\revB{
One of the ``killer" applications of Datalog has been in program analysis because it often require solving constraint systems with complex and recursive inter-dependencies, for example, call graph construction or conditional constant propagation~\cite{dl,souffle1,flix16,bdd,muZ}.
Figure~\ref{fig:tc}(a) shows an example Datalog program representing a context-sensitive pointer analysis benchmark (CSPA), originally introduced in the Graspan engine\cite{graspan}. Figure~\ref{fig:tc}(b) shows the EDB and IDB database schema with dependencies between relations illustrated with a solid black line. 
Figure~\ref{fig:tc}(c) shows an example subquery that will be generated at each iteration with the Semi-Naive bottom-up evaluation technique,
for the set of rules defining the \texttt{VAlias} predicate, highlighted in blue in Figure~\ref{fig:tc}(a).} $\delta$ represents ``delta" relations and $\star$ represents ``derived" relations. 
\subsection{Metaprogramming}
\emph{ Metaprogramming}\label{sec:background:mp} is a programming technique in which programs can treat other programs as their data, sometimes as the input (e.g., analytical macros) or as the output (e.g., code generation). \emph{Multi-Stage Programming} (MSP)~\cite{gentlemsp} is a variant of metaprogramming in which compilation happens in a series of phases, starting at compile time and potentially continuing into multiple stages at runtime.

The MSP literature breaks staging into two fundamental operations, \emph{quotes} and \emph{splices}~\cite{gentlemsp}. Quotes delay evaluation of an expression to a later phase, generally referred to as \emph{staging} an expression. Splices compose and evaluate expressions, generating either expressions or another quote.

\revC{In the Scala compiler, the construction of a quote serializes the elaborated code after type-checking into an interchange format called TASTy, which represents the typed abstract syntax tree (AST) and is stored alongside Java classfiles. Splicing deserializes quotes and executes the code if at the correct stage, determined by the difference in the number of splice scopes and quote scopes in which the splice is embedded. 
If there are more splices than quotes, then the code is run at compile-time (i.e., a macro); if the number of splices and quotes are equal, the code is compiled and run as usual; if there are more quotes than splices the code is \emph{staged}~\cite{pumm}.
% for every quote for which there is not a splice, execution is deferred to a later stage. Code generation can thus occur at compile-time, at the start of a Datalog query, or repeatedly during the execution of a Datalog query. 
%The result of staging an expression is a transformed, often partially evaluated representation of the input program that can be stored and run later. }
}

% MSP is both \emph{generative}, meaning it can generate code, and \emph{analytical}, meaning code can be analyzed and decomposed into sub-components. 
MSP enables code to be specialized at the level of abstraction appropriate for the optimization, but is also \emph{principled} as it provides safety guarantees. Unlike many code-generation frameworks that concatenate strings or inject instructions at runtime, it is not possible to generate code at runtime that is unsound. Staged code is parsed, type-checked, and partially compiled 
% into an intermediate representation by the language compiler 
during the initial program compilation, and can then be passed around within the program as a first-class value. MSP also enables \emph{deoptimization}, necessary when the assumptions that the optimization was based upon have changed and the optimization may no longer show speedup,  through deferring control flow easily between generated and existing code.

%% file: 03-arch/main-v0.tex
\begin{figure}
\centering
  \includegraphics[trim={1.8cm 20.9cm 6.1cm 2.2cm}, width=\linewidth,clip]{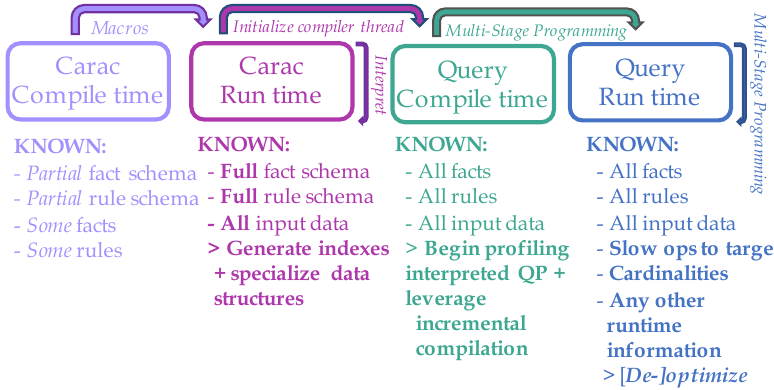}
  \vspace{-.8cm}
  \caption{Stages information becomes available}\label{fig:stages}
  \vspace{-0.5cm}
\end{figure}

\arch{} is a \emph{staged} approach: the Datalog program passes through a series of partial evaluations that continue past the start of the program runtime, re-generating code as needed. The fundamental performance benefit to code generation is \emph{specialization}, which enables avoiding as many expensive data-dependent decisions as possible while applying the best execution strategy given the available information. Without the ability to make runtime decisions, query optimizers must rely on sometimes inaccurate estimations for cardinalities or cost models~\cite{costmodels}, employ an entirely separate profiling or ``learning" stage, or miss out on certain optimizations entirely because the runtime cost is too high. 
\revA{Unfortunately, many optimizations that tune internal data structures or regenerate code have a high overhead, so shifting all optimization to runtime can be prohibitively expensive.} Staging allows for \emph{gradual} specialization, so that optimization decisions can be made or previous ones revised as soon as new information, for example fact relation cardinality or rule schema, becomes available. \revA{Rule definitions dictate the IDB schema, which contains information like relation dependencies and potential join keys.
}\revB{For example, the schema for the CSPA program is illustrated in Figure~\ref{fig:tc}(b). The dotted blue lines represent the locations of shared variables within Datalog rules, which will become join keys in the generated queries.} \revA{Knowing the join keys of queries ahead-of-time enables optimization decisions, for example the Datalog engine can automatically decide to build an index on a column where there is a join key so the index can be built incrementally before execution begins.} 

% \TODO{remove: One of the ``killer" applications of Datalog has been program analysis 
%proposed as a concise, maintainable alternative for expressing static analyses 
% ~\cite{dl,souffle1,flix16,bdd,muZ}. 
% Static analyses often require solving constraint systems with complex inter-dependencies, for example, call graph construction and conditional constant propagation. 
% Because Datalog is based on logic programming, it can express complex interdependencies and deeply nested mutual recursion with simple logic statements. 
% Static analyses that can take hundreds of lines of code to express in traditional imperative programming languages can be expressed in just a few lines of Datalog~\cite{bdd}.
% } 
The nature of program analysis is that the information required to execute and optimize efficiently is not available at a single point in time.
The diagram in Figure~\ref{fig:stages} shows potential stages within the lifecycle of a Datalog query and at which stage information becomes available.
%Because the information arrives in stages, it follows that the execution engine should be \emph{staged}. 
As an example, consider an analytical data management system that ingests user-defined functions (UDFs) and performs static analysis to identify optimization opportunities\cite{cidr}. The analytical system will have predefined what analyses it will perform, so \revA{at compile-time} it will have access to the Datalog rule and fact schema, plus any other built-in fact instances representing invariants. If the UDF is registered with the system ahead-of-time, then the facts generated from the UDF code will become available for analysis. \revA{Having access to this information ahead-of-time will enable the Datalog engine to generate an efficient initial query plan or precompile certain subqueries, avoiding expensive compilation overheads at runtime.}
%At the moment the UDF is registered with the analytical system, the UDF program itself may have just been written, but it is likely that any libraries called into by the snippet are already defined and could have been pre-ingested at an earlier point in time. Alternatively,  
Then, once the developer wants to deploy the UDF in an analytical pipeline, knowledge of the adjacent operators becomes available so only then can inter-operator analysis queries execute. For subsequent runs of that analysis, the optimizer should be able to use relevant information from the previous runs, \revA{for example statistics about the input fact data that would inform the initial join-ordering}. Alternatively, if the UDF is not registered ahead-of-time, then only the rule schema will be available to inform the query optimization ahead-of-time. Further, if the analytical system ingests custom analyses, then neither the facts nor the rule schema will be available before query execution begins and runtime optimization is the only option. Staging, via phases of compile and runtime code generation, allows the optimizer the freedom to leverage concrete information as it becomes available \revA{to turn as many expensive optimizations as possible into offline costs while still seeing runtime performance benefits from specialization}.

%% file: 04-optimizations/main-v5.tex
% \edit{In this section, we present an example runtime optimization that is applied within \system{} to accelerate the execution of metaprogramming queries.} \TODO{Could argue that this optimization could also accelerate user pipelines that have complex joins but then I would expect an experiment to support that statement, which I likely will not have time to do.} 
The order of atoms within a Datalog rule definition does not affect Datalog semantics. Thus rules may be rewritten with different orders of atoms without changing the result of the program. The order of atoms can, however, have a significant effect on the performance of the program. Because of this, we chose to target rearranging the order of atoms within a rule to illustrate the optimization potential of \arch{}. In the following section, we explain why reordering atoms can lead to performance gain and present the optimization algorithm applied.

As can be seen \revB{from the query illustrated in Figure~\ref{fig:tc}(c),} for a rule definition with $n$ atoms in the body that share at least one variable with another atom, evaluation requires a union of $n$ queries \revA{where} each contains an $n$-way join of $n-1$ derived relations and $1$ delta relation.
There are a factorial number of possible left-deep join orders for each sub-query, and it is well known that the order in which join operations are executed can profoundly affect the execution time~\cite{joinorder}. When doing  $n$-way joins, there is the potential for the intermediate relations to blow up in size significantly, only to be reduced in size in the final join. A more performant execution plan could avoid a cardinality blow-up by carefully selecting the order of atoms so that the sub-joins produce intermediate relations with smaller cardinalities. However, \revA{existing cardinality estimation techniques\cite{costmodels}} cannot easily predict ahead of time what the cardinality of the relations will be because the result of the previous loop is fed into the next iteration, and even within the same iteration, we may be 10 relations deep into an $n$-way join and predicting the size of the join sub-tree gets complex. %It is nontrivial to predict the behavior of delta relations as well, as they may start small and drop quickly, or do the reverse.

% \pagebreak

\revA{To avoid the complexity and inaccuracy of multi-iteration cardinality estimation, \system{} uses \arch{} to dynamically regenerate query plans using runtime information. Concrete instances of relations are plugged directly into a join reordering algorithm at the last possible moment, reducing the complexity and inaccuracy of the cost model. \arch{}, applied within a different system that uses an alternative cost model, would tune the reordering algorithm to the different costs of each operation.}
% NOTE: the point is that we still have a cost model but we take the guesswork out of it

\revA{Three input parameters \system{} takes into account to select the join ordering are input relation cardinality, index selection, and the selectivity of the join condition.

\revC{\textbf{Input relation cardinality.}} The cardinalities, including those of any already-computed intermediate relations, are available to be read at the time the optimization is applied. %Thus, the most up-to-date information about relation cardinality is always available to inform 2-way join order, for example selecting the build or probe side of the join.

\revC{\textbf{Index selection.}} As each rule is defined, \system{} builds an index for the columns of each relation that has a constant or variable in the body of the corresponding rule.
%. When there is a constant constraint, e.g. there is a constant term in the body of an atom, a primary key index will be built on that column because \system{} always pushes constant predicates before column predicates. When there is a variable term that is shared with another atom in the body of the rule, a secondary index is added for those relations.
There is a rich field of study on auto-index selection\cite{indexdl} to cut down the number of indices but the approach taken by \system{} is to build one index per filter or join predicate.

\revC{\textbf{Selectivity.}} \system{} uses a constant reduction factor for each additional constraint with the assumption that each condition is statistically independent. It is of course possible to maintain more detailed statistics, for example online histograms\cite{cow}, but \system{} aims to keep the reordering algorithm as lightweight as possible to avoid runtime overhead.
}
The optimization can be first applied as soon as the Datalog rule schema is available. \revB{We take the CSPA Datalog program illustrated in Figure~\ref{fig:tc}, applied to facts generated from the Apache httpd source code as a running example, which contains about 1.5 million input facts. For brevity, we focus specifically on the second rule definition for \texttt{VAlias}, which contains three atoms and, as can be seen from Figure~\ref{fig:tc}(c), three queries with a 3-way join between the \texttt{VaFlow}, \texttt{MAlias}, and \texttt{VaFlow} relations which are themselves IDB relations (so their cardinalities will not be statically known after the first iteration). 
Consider the subquery where we read one $VaFlow_\delta$ relation from the delta database, one $VaFlow_\star$ relation from the derived database, and the $MAlias_\star$ relation from the derived database. There are no conditions between $VaFlow_\star$ and $VaFlow_\delta$ so the result will be the cartesian product. Looking at the very first iteration, where the cardinalities are: $\left|VaFlow_{\delta}\right| = 541096$, $\left|VaFlow_{\star}\right| = 903752$, $\left|MAlias_{\star}\right| = 541096$, the intermediate relation of the subquery $\left|VaFlow_\star\bowtie VaFlow_\delta\right| = 816\,767\,677\,504$, which is about 6534GB as each tuple contains 2 32-bit integers. Alternatively, if we reorder the join, we can have $\left|MAlias_\star\bowtie VaFlow_\star\right| = 903752$. However, if we peek at execution during the 7th iteration, the input cardinalities are as follows:
$\left|VaFlow_{\delta}\right| = 0$, $\left|VaFlow_{\star}\right| = 1362950$, $\left|MAlias_{\star}\right| = 79514436$, then not applying the empty relation to the first join would be costly, so \system{} would select $VaFlow_\delta \times MAlias_\star$. In the case of the CSPA example program, there are at maximum 3-way joins, but other Datalog programs with longer rules will suffer from even more variability (e.g. the Inverse Function analysis described in Section~\ref{sec:eval:aot} has a 9-atom rule), especially ones without cartesian products as the selectivity estimation can attempt to prevent them even without cardinality information.

%, 00x larger than the maximum value representable by a 32-bit integer. Because \system{} currently uses 32-bit integers to index into arrays, there exists a current limitation that \system{} cannot support relations with larger than the largest 32-bit integer. Custom data structures or clustered indexes remain as future work, but to illustrate the example while avoiding this limitation, if we take a sample of 20,000 tuples from the httpd dataset, the input relations at the same point in execution are $VaFlow_{\delta}|541096|,\ VaFlow_{\star}|903752|,\ MAlias_{\delta}|0|,\ MAlias_{\star}:|541096|$.
%. Because the \texttt{MAlias} delta relation is empty, the final result will be empty and we will have had a completely avoidable materialization of about 0GB. However, it is not enough to simply stick with that ordering for the entire execution: for example, at iteration 0 the optimal join ordering using \system{}'s cost model would be to put the \texttt{VaFlow}
}

The optimizations can be at \emph{\system{} compile-time} and implemented with compile-time generative macros. Depending on what the user supplies, at compile-time the algorithm may not yet have access to the relation cardinalities, as some facts have not yet been added to the system, so the join-ordering will rely on the available facts and the selectivity heuristic. 
% The performance of query execution based on what stage facts and rules become available is presented in Section~\ref{sec:eval:aot}. 
The optimization can also be performed at \emph{query compile-time}, where the rules and cardinalities of the initial extensional database are available. If facts are only provided at runtime then this will happen at the start of query execution. 
Further, since \arch{} enables on-the-fly recompilation of subqueries, this optimization can be repeatedly applied over the course of query evaluation to adapt to changes in the relation cardinalities.

%% file: 05-carac/main-v3.tex
\begin{figure}
\centering
    % \frame{
    \includegraphics[trim={.8cm .8cm .9cm .6cm}, width=.9\linewidth,clip]{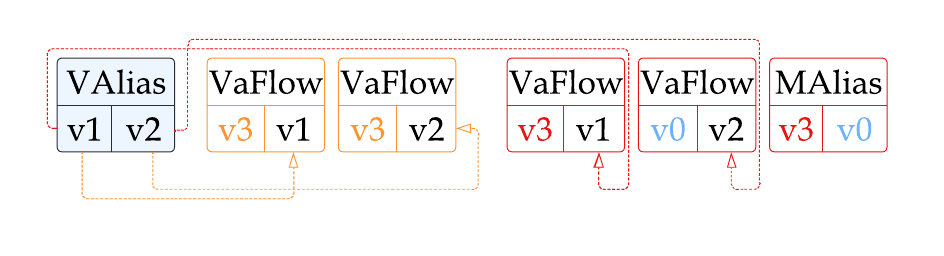}
    % }
    \vspace{-0.3cm}
    \caption{\texttt{VAlias} Datalog AST}\label{fig:ast}
\vspace{-0.3cm}
\end{figure}
\begin{figure}
\centering
    % \frame{
    \includegraphics[trim={.6cm .7cm .9cm 4cm}, width=\linewidth,clip]{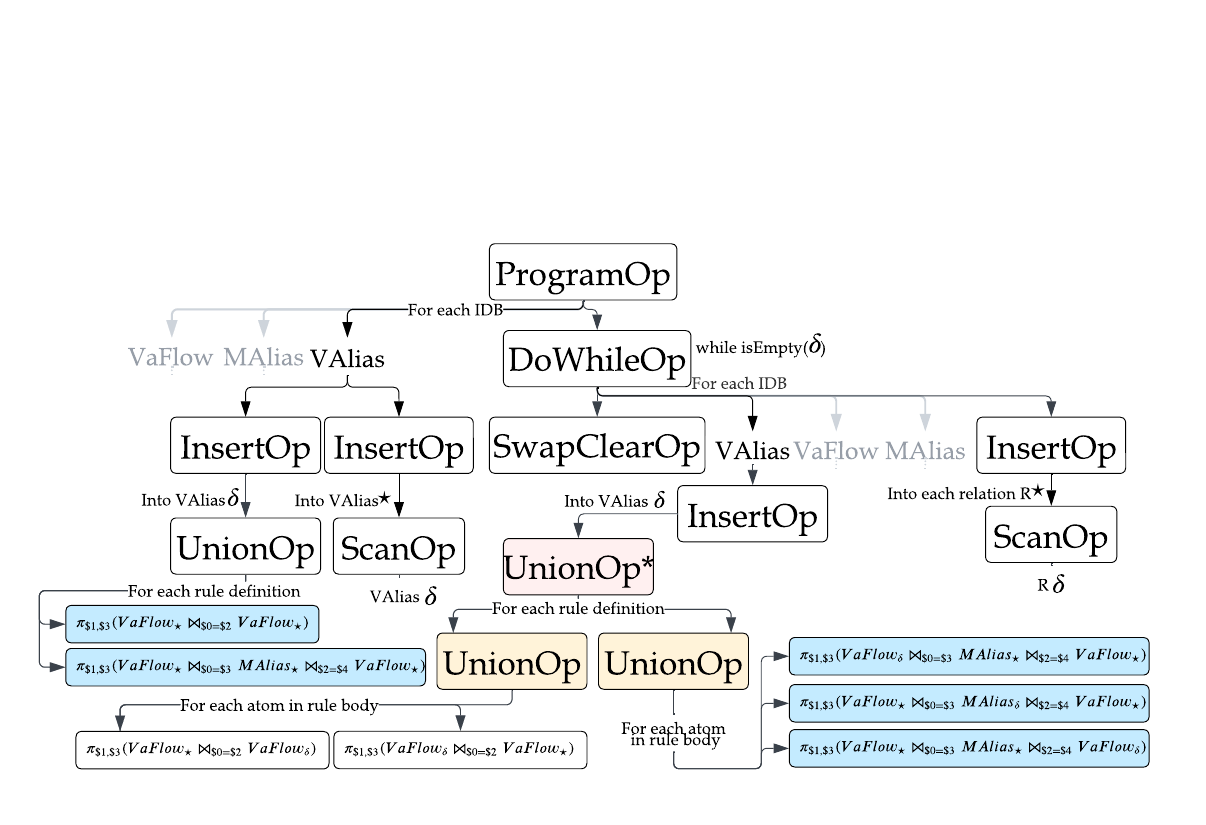}
    % }
    % \vspace{-0.2cm}
    \caption{Carac IROp Program Structure}\label{fig:tree}
\vspace{-0.5cm}
\end{figure}

In this section we provide an overview of the system architecture of \system{} and show how \system{} progressively lowers the input Datalog query into an executable. Section ~\ref{sec:sys:dsl} describes the user-facing embedded DSL, Section ~\ref{sec:sys:ee} describes the generation and optimization of the logical query plan, Section~\ref{sec:sys:compilation} discusses generation of the physical plan for the Datalog-specific operators, and finally Section ~\ref{sec:sys:storage} describes relation storage and generation of the physical plan for the relational operators. 

We chose to write \system{} in Scala because it provides \emph{Principled Metaprogramming}\cite{gentlemsp} combined with compile-time macros~\cite{nicothesis}, higher-order functions, and direct generation of JVM bytecode at runtime. 
%In the Scala compiler, construction of a quote essentially pickles the fully elaborated code after type-checking into an interchange format called TASTy~\cite{pumm}, which represents the typed abstract syntax tree (AST) and is stored alongside Java classfiles. Splices unpickle quotes and potentially execute the code if at the correct stage %. The \texttt{staging} library provides using an instance of the Scala 3 compiler available at runtime. Generated code is run at the phase determined by the difference in the number of splice scopes and quote scopes in which it is embedded: if there are more splices than quotes, then the code is run at compile-time (i.e., a macro); if the number of splices and quotes are equal, the code is compiled and run as usual; if there are more quotes than splices then the code is \emph{staged}. Each subsequent extra quote is another stage. 
Multi-Stage Programming language constructs allow us to generate code at compile-time, at the start of a Datalog query, or repeatedly during the execution of a Datalog query. The result of \emph{staging} an expression is a transformed, often partially evaluated representation of the input program that can be stored and run later. In \system{}, this can happen at different levels: the internal intermediate representation, stored higher-order functions, quotes containing typed ASTs of Scala code, or direct generation JVM bytecode. %Section~\ref{sec:sys:compilation} presents details on the different staging mechanisms. 
% \edit{We chose to run our \system{} queries over Scala programs because of the popularity of the language among data scientists and the convenient intermediate representation, TASTy, but the principles illustrated extend \system{} to extracting facts from any intermediate representation, for example, LLVM-IR or Java bytecode as is done by the Soot framework in the Doop analysis framework~\cite{doop}.}
\subsection{\system{} DSL}\label{sec:sys:dsl}

\system{} supports a deep embedding of a Datalog-based DSL into the Scala programming language extended with stratified negation and aggregation. Similarly to Flix~\cite{flix16}, \system{} supports first-class Datalog constraints for interoperability between general-purpose programming and the declarative Datalog programs. \revB{An example Datalog program in \system{} is shown in Figure~\ref{fig:tc}.}
%A simple transitive closure Datalog program expressed in \system{} can be seen in Figure~\ref{fig:tc}. On line 0, the program environment is constructed from two components: (1) an execution engine instance, responsible for the Datalog evaluation strategy, and (2) a relational storage and execution layer, responsible for the management of the generated relations and execution of relational subqueries.% (discussed in detail in Sections~\ref{sec:sys:ee} and ~\ref{sec:sys:storage}). 

%Two facts for the relation \texttt{edge} are defined on lines 6 and 7, and two rules for \texttt{path} are defined on lines 4 and 5. 
%Each clause within a Datalog rule definition is called an \emph{atom}, e.g., \texttt{edge(x, y)} within the rule definition on line 5. 
%On the final line 8, we can call \texttt{solve} on our rule \texttt{path} to begin execution of the query and get all the facts that can be generated from the input facts using the rule \texttt{path} and any other rule or fact relations that it depends on.

%\subsubsection{Stage 0: Program Definition}~\\
\system{} facts and rules can be defined at compile-time or incrementally added at runtime. Each fact is added to the corresponding relation in the relational layer as it is defined. As rules are defined, they are added to the Datalog AST, and metadata is generated for each rule. The metadata includes the locations of variables and constants in the rule head and body, which will later be used to optimize execution, and generation of a precedence graph so that relations that rely on other relations will be calculated only after their dependencies are calculated. \revB{The AST generated from the running CSPA example, for the \texttt{ValueAlias} rules, is illustrated in Figure~\ref{fig:ast}. The CSPA program does not contain any relation aliases, but if there were, then a static rewrite pass would remove any aliases to avoid extra costly materialization.}
%For recursive dependencies, the strongly connected components are determined so they can be evaluated within the same fixed-point calculation. 

\subsection{Execution Engine}\label{sec:sys:ee}

% in the form of an intermediate representation comprising relational algebra operators, control flow statements, and relation management operations. 
In this section, 
The main challenges that inform the design of \system{} are discussed, including the generation and optimization of the logical query plan and how \system{}’s JIT switches from interpreting the logical query plan to generating code for the physical query plan.
% the main challenges that inform the design of \system{} are discussed, include managing state, switching execution modes part-way through execution of a single query, and the tradeoff between the performance, granularity, and expressivity of generated code.

% \vspace{1em}
\subsubsection{Logical Query Plan Generation}\label{sec:ee:log}
 
\system{} uses a Futamura Projection~\cite{futamura} to generate an imperative program specialized to the input Datalog program, similar to Soufflé's Relational Algebra Machine~\cite{souffle1}. \revB{The Datalog AST, illustrated in Figure~\ref{fig:ast}}, is visited and the Semi-Naive evaluation strategy is used to partially evaluate the Datalog program into a series of operations called IROps, illustrated in Figure~\ref{fig:tree}.  \revC{Operations with multiple child nodes are executed from left to right. The IROp intermediate representation consists of the following operations: relational algebra expressions, e.g. Select-Project-Join ($\sigma \pi \bowtie$) in blue, Union in yellow and pink, and set difference (not illustrated as it is inlined into the join operation); control flow operations, e.g. DoWhile, and relation management operations, e.g. InsertOp, ScanOp, and SwapClearOp. To avoid expensive set difference operations, and to facilitate asynchronous compilation and parallelization, the database storing the delta relations is split into a read-only ``known" database, that stores tuples discovered in the previous iteration, and a write-only ``new" delta database where tuples discovered in the current iteration are written. At the start of each iteration, SwapClearOp swaps the read-only delta database with the write-only ``new" delta database, and clears the relations that will become the write-only delta relations in the next iteration.}
\system{}'s IROps are represented with generalized algebraic data types (GADTs)\cite{gadt}, so they can be either interpreted or compiled into a function that returns the correct type. 

While the generated IROp program is imperative, we consider it the logical query plan for both the Datalog-specific operators and the traditional relational algebra operators. When \system{} is in interpretation mode, there is no further partial evaluation and the interpreter visits this IROp tree.

%^All IROp programs have a similar core structure, shown in Figure~\ref{fig:tree}. 

%One of the optimizations applied is loop unrolling, so the IROp program trees will grow proportionally to the number of relations. For clarity, we have included only the subtrees for the \texttt{VAlias} relation illustrated in Figure~\ref{fig:tc}, but other relations (notated in grey) will be represented with an analogous structure.

%In the first stage after execution begins, a series of static rewrite rules are applied to the AST to perform optimizations like alias elimination. After the AST transformation phase, the AST is visited to transform the declarative Datalog constraints into an   Static optimizations like predicate pushdown and for-each-relation loop unrolling are performed in this phase.

\subsubsection{Compilation Granularity}
\revC{Any runtime optimization must manage the classic tradeoff between the freshness of the data used to make optimization decisions and the overheads of reoptimization: ideally \system{} would apply the reordering optimization every time a join is executed to ensure the optimal order, however, regenerating plans include overhead that may be great enough to outweigh the benefits of a more optimal join order.

As a result, we have designed \system{} as a just-in-time optimizing compiler. How frequently new optimizations are applied and a new query plan is generated is called \emph{compilation granularity}.
} When in JIT mode, \system{} will begin execution by interpretation at the root node of the tree, and can switch to compilation at any node in the IROp program tree. Nodes that appear higher in the tree have fewer instances, so compiling at a higher granularity will incur fewer compilations, but the code being compiled will be larger, and the information available to make optimization decisions may be outdated compared to compiling at a lower granularity. \revC{For example, when compiling at the pink UnionOp* in Figure~\ref{fig:tree},} the optimization will be applied once for every relation, every iteration of the Semi-Naive algorithm. The cardinalities used will be from the Derived database, i.e. the state of the relations at the start of the outer program loop. At a lower granularity, like the blue $\sigma \pi \bowtie$, the optimization will occur before every \emph{n}-way join. The relation cardinalities at the lower granularity will be the most accurate, as they will take into account the exact cardinality of the delta relations, but the compilation will happen more frequently.

To avoid unnecessary compilations, \system{} applies a ``freshness" test before recompiling on higher-overhead compilation targets to check that the cardinalities have changed, relative to other relations, above a tunable threshold. For the case of atom reordering, it is possible to push optimization entirely to runtime for the IRGenerator compilation target described in Section~\ref{sec:sys:compilation}. However, this limits the ability of \system{} to combine it with other lower-level optimizations enabled by code generation, e.g. just-in-time data structure specialization. 

When \system{} has begun compilation, it can either block waiting for the generated code to be callable or compile asynchronously and continue interpretation, switching to the compiled code only when it's ready. Asynchronous compilation is intended for higher-granularity operations: for example, a union operation of \emph{n} subqueries will begin compilation, then start interpreting the first sub-query. After each subquery is executed, the interpreter will check if the compilation is ready - if so, it will call into the compiled code such that it begins at the exact spot the interpreter left off. \revC{In the IROp AST for the CSDA program illustrated in Figure~\ref{fig:tree}, compiling at the pink UnionOp* will require one compilation per iteration per relation. Compiling at the yellow UnionOps will require one compilation per iteration per rule definition. Because the relation cardinalities will not change between UnionOp* and UnionOp, the difference between starting compilation at the lower granularity is that more compilations will occur, but the interpreter will check for a ready compilation after every rule definition is processed, while the higher granularity will have fewer compilations but only be able to check for a ready compilation after each relation is processed.
}

The goal is for \system{} in JIT mode to not perform substantially worse than pure interpretation since even if compilation takes unexpectedly long, the interpreter can continue making progress evaluating the Datalog program while compilation happens on a different thread. However, this means that if the compilation takes longer than the time to finish interpretation of the sub-tree, then the generated code may never be used.% within that iteration. %In the next iteration, if the data has changed enough, recompilation may be triggered and the code will be unused.

% \vspace{1em}
\subsubsection{Safe Points}\label{sec:ee:comp2}%~\\
``Safe points" refer to locations within the execution of a program where an optimizing JIT compiler can safely switch between interpretation and compilation. When the \system{} JIT reaches a node to compile, it can compile that node in ``full" or ``snippet". When compiling in full, the node itself and its entire subtree are compiled into a single function. The interpreter will continue to visit nodes until it reaches the node of the correct granularity and then will switch entirely to the compiled code, only switching back once the entire subtree has been evaluated. The next time the interpreter reaches that node, it can use the existing compiled code, recompile it into a new function, or revert back to interpretation. 

Alternatively, \system{} can compile in ``snippet" when using the quotes compilation target, which compiles only the body of the targeted node, no children, and then control flow is deferred back to the interpreter from the compiled code by splicing continuations into the generated code. Compiling only snippets enables continuous checks for novel optimizations or \emph{deoptimization}, which extends the staging into however many levels as needed. Additionally, snippet compilation allows the absolute minimum amount of code to be generated to mitigate some overhead costs of compilation.

All program state is either returned from functions in the form of generated facts or stored in the relational storage layer. This property is why any IROp node can be considered a ``safe point" to switch evaluation modes in \system{} and the stack does not have to be saved or copied.

% \vspace{1em}
\subsection{Compilation Targets}\label{sec:sys:compilation}%~\\
\system{} supports four different targets for compiling IROps into the physical query plan. The targets range from highly expressive and flexible, allowing for arbitrary runtime code generation and continuous re-optimization, to performant but limited or unsafe code generation.
%, some targeted at expressiveness and type-safety which the tradeoffs will be discussed in Section~\ref{sec:exp}. 

% \vspace{1em}
\subsubsection{Multi-Stage Programming Quotes \& Splices}\label{sec:ee:q}

The most powerful, expressive, and safe mechanism for \system{} to generate code at runtime is to compile IROps into the Scala compiler's \emph{Quote} and \emph{Splice} constructs, which implement classic Multi-Stage Programming concepts~\cite{gentlemsp} \revC{described in Section~\ref{sec:background:mp}}. Quotes delay execution for a later stage and can be thought of as code generation, while splices trigger the execution of a quote. Quotes can contain other quote instances, enabling multiple stages of code generation. Nodes in the IROp program tree are compiled into quotes containing a single function definition at the outer-most level so that when it is time for the code to be executed, it can be called exactly like any other function. Once an operation is compiled, calling the function has no more overhead than an ahead-of-time compiled function. Quotes allow for de-optimizing if necessary, reverting back to interpretation, or further regenerating code. 

\revC{For example, when we compile the CSPA example at the yellow UnionOp illustrated in Figure~\ref{fig:tree}, every time we reach a tree node of that type for the first time, we apply the optimizations and begin constructing a quote. This quote contains a single function that, when called, executes either a union of 2 $\sigma\pi\bowtie$ (when visiting the left-most UnionOp) or 3 $\sigma\pi\bowtie$ (when visiting the rightmost UnionOp). If compiling synchronously, we store the quote and then execute it. If compiling asynchronously, we immediately start interpreting the first child and launch a subthread to compile the $1\rightarrow n$ other child nodes. Every time we return to that node, we check for the quote, and if it's ready, we call the quote; otherwise, we continue interpretation. When compiling in ``snippet" mode, we call the quote and pass the quote function a continuation, so the generated code can call back into the interpreter. If not in snippet mode, execution continues only inside of the quote, meaning we will never visit a subtree of that node unless the freshness test is failed and the quote is regenerated.}

\begin{figure}
\centering
% \frame{
  \includegraphics[trim={1.2cm 1cm 1cm 2.6cm}, width=\linewidth,clip]{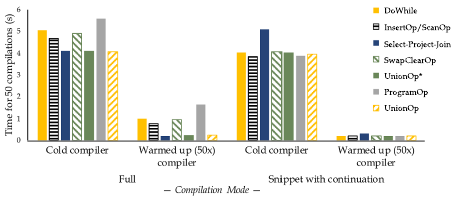}
  % }
  \vspace{-0.8cm}
  \caption{Execution time of code generation}\label{fig:dotty}
  \vspace{-0.5cm}
\end{figure}

Because quotes are type-checked by the Scala compiler at \system{}'s compile-time, they are both safe and the most powerful mechanism for code generation, as they allow for unlimited functionality within the generated code. However, the safety and expressiveness come at a price: compiling quotes into bytecode requires invoking the Scala compiler at runtime, which sometimes incurs significant overhead costs.
Figure~\ref{fig:dotty} compares the quote generation time depending on if the compiler is ``warm", i.e. compilation has happened at least 50x before measurement, and ``cold" meaning the compiler is instantiated immediately before compilation. The legend displays at which granularity (illustrated in ~Figure~\ref{fig:tree}) compilation happens. The left-hand side shows ``Full", meaning compilation of the entire subtree, and the right-hand side shows ``Snippet", which compiles only the operator and splices continuations to revert to the interpreter after the operator completes execution.  The safety properties are especially important when considering Datalog language extensions that allow for arbitrary code execution, similar to a UDF in a traditional data management system. The quotes and splices backend enables safer expressivity of the input Datalog program because it prevents unsafe user code from executing within \system{}.

% \vspace{1em}
\subsubsection{Bytecode} Alternatively, \system{} supports direct generation of JVM bytecode from IROps, using the experimental JEP Class-File API Preview~\cite{jep}. The bytecode generator can be invoked at runtime, but unlike Quotes, it does not generate code that can revert to interpretation once compiled. However, if the interpreter reaches a node that has been compiled into bytecode it can choose to throw out the previously generated bytecode and regenerate.

Direct-to-bytecode generation is, in theory, as expressive as quote generation as it also allows for arbitrary code execution. However, manually generating bytecode instructions is error-prone and requires a much higher development cost than the Quotes API, which only requires writing Scala code as one would for regular, non-staged code. Further, because the generated bytecode is interpreted directly by the JVM, there are few safety guarantees. Bypassing the safety checks of invoking a full compiler leads to lower overheads but at the cost of losing the ergonomics and safety of writing in a type-safe programming language.

% \vspace{1em}
\subsubsection{Lambda} \system{} supports a lambda-based backend that, at runtime, stitches together higher-order functions that are precompiled at \system{}'s compile-time. The Lambda backend is a more limited compilation target as \system{} can only choose between predefined procedures. However, the cost of invoking the Scala compiler at runtime can be avoided while maintaining the safety guarantees lacking from bytecode generation. 
%This approach is analogous to LLVM-IR generating analytical systems that inject jump instructions between pre-compiled paths. 
The performance of the generated lambdas benefits from the ability of the JVM to aggressively inline these smaller functions to avoid overhead costs of function calls.

% \vspace{1em}
\subsubsection{IROp-Generation} Lastly, \system{} can regenerate IROps on the fly and invoke the interpreter to execute the new IROps when available, called ``IRGenerator". IROps are the most limited target, as they only allow for rearranging operations within the existing IROp program structure. For the case of the optimization described in Section~\ref{sec:opt}, it suffices but would not scale to other optimizations that are applied at a lower level of abstraction than the IROps.  While the program is still staged, as the IROps are a partially evaluated representation of the input Datalog program, the IRGenerator target does not need to invoke a separate compilation phase so the overhead is limited to the cost of sorting and rearranging the subqueries at runtime.  %However, it does benefit from the lowest overhead of \system{}'s backends because the online cost is limited to constructing new in-memory IROp nodes within the IR Tree.

The tradeoffs between the flexibility and performance of the different compilation targets are evaluated in Section~\ref{sec:eval}. The choice of backend is left as a user parameter because \system{} cannot anticipate if a user may wish to prioritize performance over safety (and would therefore use the bytecode target), expressibility of the Datalog variant over performance (quotes and splices), safety over expressivity (lambda), or are otherwise limited due to the execution environment (for example, using Graal Native-Image one cannot use runtime classloading and could therefore only use lambda and IROp-generation), while running on the JVM enables all backends.

% \vspace{1em}
\subsection{Physical Relational Operator Layer}\label{sec:sys:storage}

\system{}'s execution layer is built on top of a pluggable ``relational layer" to store input and intermediate relations and plan and execute basic relational algebra operations. Each rule has two types of intermediate relations: a \emph{Derived} database (all discovered facts) and a \emph{Delta} (newly discovered) database. 
% The Derived database stores all the facts discovered so far throughout a Datalog program evaluation. The Delta database stores only the new facts derived in the previous iteration of the main program loop. 
The Delta database is further split into \emph{Known} and \emph{New} databases to isolate the reading of previously derived facts within a subquery and writing of newly derived facts in the same query. This split between read-only and write-only relations, which are swapped at the end of each iteration, is what enables flexibility in \system{}'s safe points as well as parallelization of queries at the Union or Union*.

The relational layer API provides read and write access to the relations, as well as \texttt{clear}, \texttt{swap}, and \texttt{diff} operations. \texttt{swap} and \texttt{clear} serve to efficiently implement copying all the newly derived facts into the existing derived facts at the end of every loop iteration, and \texttt{diff} determines if the loop should terminate. Finally, the relational layer API provides relational operators that make up the sub-queries: \texttt{select}, \texttt{project}, \texttt{join}, and \texttt{union}. Typical relational algebra optimization and compilation can be performed at this level.

% \edit{Because these requirements are universal to many analytical systems, \arch{} does not need to be implemented as a standalone system and can be integrated into existing analytical systems. This unification of program analysis queries with user-provided data pipelines enables (1) system resources to be fully utilized and (2) program analysis queries to be run just-in-time to the runtime of the data pipeline so the analytical system can take advantage of both runtime data information and strong, programming language-derived properties of input programs to apply optimizations that would not be possible with only runtime or only compile-time analysis. 
So far, \system{} has been integrated with a typical push-based and a pull-based engine using in-memory Scala collections. 
%It has been shown that depending on the characteristics of the query being executed, there are different advantages to each engine~\cite{pushpull}. 
% Alternatively, the \texttt{DistributedStorageManager} relies on Spark to provide fact storage and relational algebra operators.

% While Datalog and most data management systems operate over relations, the goal of \system{} is to operate over representations of programs. The Scala 3 compiler generates a high-level interchange format, named TASTy (for Typed Abstract Syntax Trees), that contains the compiled program's syntactic structure, types, positions, and even documentation. Quotes are in-memory versions of TASTy trees. During compilation, the Scala 3 compiler serializes these trees and saves them to a \texttt{.tasty} file that is stored with the generated \texttt{classfiles}. To operate over programs, \system{} can ingest TASTy files directly and automatically generate the facts required to run certain predefined analyses similar to how the Doop ~\cite{doop} framework generates facts from the JVM bytecode. Currently, \system{} targets Scala programs, but future work could be extended to other languages. For this to happen, the generated facts and the analysis rules would need to be updated for the language semantics but the core execution would not change.

%% file: 06-eval/main-v3.tex
% BATCHSIZE 10, on 46
To evaluate \system{} with a range of query structures, we construct a set of queries to benchmark that include \revA{macrobenchmarks, i.e. queries that are representative of program analyses, and microbenchmarks, i.e., more general recursive queries}. In Section~\ref{sec:eval:queries}, we discuss the details of these queries and why they were chosen. In Section~\ref{sec:eval:opt}, we evaluate \system{}'s runtime-optimized query performance compared to interpreted queries. In Section~\ref{sec:eval:aot} we evaluate \system{}'s compile-time-optimized query performance compared to runtime-optimized.
%, as well as the combination of compile-time with runtime optimization. 
In Section~\ref{sec:eval:sota}, we compare \system{}'s runtime with a current state-of-the-art Datalog execution engine, Soufflé. 
% \edit{Lastly, in Section~\ref{sec:eval:pipeline} we compare execution of a simple data pipeline that includes two user-defined executables interweaved with traditional relational algebra operators.}

Experiments are run on an Intel(R) Xeon(R) Gold 5118 CPU @ 2.30GHz (2 x 12-core) with 395GB RAM, on 
Ubuntu 22.04 LTS with Linux kernel 6.5.0-17-generic %42
% Ubuntu 18.04.5 LTS with Linux kernel 4.15.0-151-generic %44
and Scala 3.3.1-RC4. The JVM used is Java HotSpot(TM) 64-Bit Server VM Oracle GraalVM 17.0.10+11.1 (build 17.0.10+11-LTS-jvmci-23.0-b27, mixed mode, sharing), with the flags \texttt{-Xmx396G -XX:+ParallelGC}. We use Java Benchmarking Harness (JMH) ~\cite{jmh} v1.37 with \texttt{-i 3 -wi 3 -r 10 -w 10}.
%with 3 warm-up iterations, averaged over 3 iterations, and a minimum execution time of 10 seconds per benchmark unless otherwise noted.

\subsection{Carac Query Benchmarks}\label{sec:eval:queries}

\revA{\noindent \textbf{Macrobenchmarks.}} To select representative queries for our evaluation, we first consider a custom Datalog-based static analysis query to identify potential ``wasted" work in the form of adjacent inverse functions. 
% Note that most C optimizing compilers will not identify and reverse inverse operations. 
For example, for analyzing a C program, we can add the fact \texttt{inverse(itoa, atoi)} for integer-to-string conversions (but not the reverse, since \texttt{atoi} is not a total function on the domain of strings). Where possible, calls to entire symmetric serialization libraries can be declared as inverse, e.g., \texttt{inverse(to\_json, from\_json)}. 

% revisit our motivating example from Section~\ref{sec:motex} and expand it to operate over Scala code. 
For the input data to this query, we select an input Scala program to analyze that implements and utilizes a custom linked list library in Scala that supports a \texttt{serialize} and \texttt{deserialize} function. The entry point to our input Scala program instantiates the custom linked list, operates over the list's contents, serializes the list, does some computation, then deserializes the list and returns the result. The input Scala program, which we named ``SListLib", is a minimized representation of a typical off-the-shelf utility within an analytical pipeline that does a bit of ``wasted" work that can be optimized (in this case, extra serialization).
%It has been shown in previous work \cite{cidr} that eliding serialization overhead within similar pipeline utilities can lead to a 22x speedup in end-to-end latency. 
\system{} automatically derives facts from our input Scala program by querying over the TASTy files using TASTy Query~\cite{tastyquery}. The input Scala program is around 200 lines of Scala code. % (~800 facts) and does not require the use of the Scala standard library.
% \TODO{Include or not?: Although there are interesting optimizations that can leverage the fact that both \system{} and user-supplied programs are in Scala (for example, applying \system{} to query over its own source code to identify internal inefficiencies, or inefficiencies between \system{} and any ingested utilities), we purposefully do not include those optimizations in the evaluation to emphasize that the concepts presented in \arch{} can be applied to query over input programs of any language.}

For the input \system{} query, we extend a points-to analysis query with rules that define when values could be considered equivalent given adjacent functions that can revert data transformations.
% based on those in Figure~\ref{fig:motex}(d), further elaborated to handle edge cases and updated to operate over our TASTy-generated facts. 
We then add the knowledge of which functions can ``undo" each other given the correct arguments with the fact containing the name of the list library serialization functions: \begin{verbatim}    InvFuns("deserialize", "serialize")\end{verbatim}
\revA{We evaluate \system{} with our inverse function analysis and a typical ``points-to" analysis adapted from the Doop\cite{doop} execution framework's implementation of Andersen's (context-free and flow-insensitive) points-to analysis\cite{spa} on the facts generated from the SListLib program. We also use our running context-sensitive pointer analysis example, CSPA, and a context-sensitive data flow analysis (CSDA) on the \texttt{httpd} dataset, also adapted from Graspan.}

\revA{\noindent \textbf{Microbenchmarks.}} We also include a few general, shorter-running Datalog programs: an Ackermann function\cite{ackermann}, a Fibonacci program, and a prime-numbers program. The purpose of these quicker programs is to illustrate how the complexity of the \system{} query affects the ability of the runtime optimizations to speed up the query. When talking about runtime optimizations, the longer the program the more time there is to calculate and apply optimizations without introducing disproportionate overhead. Shorter, faster-running programs are much less forgiving than longer, more complex program analysis queries. These \revA{microbenchmarks} identify the point at which the queries become so short-running that the online optimization no longer shows meaningful speedup.
% \TODO{Possibly remove the non-program analysis experiments, but I think they are worthwhile to show that (1) the optimizations are not so expensive that they only work for super-long running queries, and (2) we explored the input query space until we found the "breaking" case where the optimizations no longer show meaningful speedup.}

%We chose to base this ``reversible" analysis on a context-free points-to analysis, but in the future, we would like to extend it to context-full analyses to allow for more aggressive optimizations than are currently enabled.
\subsection{Online Optimization Speedup}\label{sec:eval:opt}
\input{06-eval/figs}

To evaluate the effectiveness of \system{}'s runtime optimization and staging mechanisms, we consider staged queries, executed with the online optimization described in Section~\ref{sec:opt}, compared with interpreted ``baseline" queries. Because there is no ``typical" way to order Datalog atoms, we consider two formulations of our input \system{} queries \revA{approximating} the best and worse cases.

One formulation has been hand-optimized and rewritten manually by carefully stepping through execution and tracking intermediate relation cardinalities. Hand-optimizing Datalog programs is a tedious process requiring advanced knowledge of Datalog evaluation algorithms and the internal execution engine implementation details. For non-experts, finding the correct join order for complex rules has been called an ``insurmountable" challenge~\cite{souffle2}. The other formulation of the \system{} queries is written to be inefficient, simulating what happens if a naive user were to have bad luck in their order of atoms. There is no reliable way for a user to tell the difference between a good and bad ordering of atoms, so we compare the most and least efficient formulations we could identify to illustrate the speedup bounds of the optimization. As it is also nontrivial for \system{} to determine if a query is optimized ahead of time or not, we use both types of input query to evaluate \system{}'s ability to optimize programs that have room to improve without overly penalizing programs whose performance is already close to optimal.

In all experiments except when comparing with the state-of-the-art systems, \system{} executes single-threaded with only compilation occurring on a separate thread. % in ``full" mode. 
When compiling on a separate thread, \system{} can either block (``blocking") or continue interpretation until the compiled code is ready (``async"). The storage engine is a push-based execution engine that stores relations in-memory in Scala Collections. The choice of staging mechanism within \system{} is referred to as the ``backend". \revA{Indexes are implemented using a Java standard library HashMap implementation}. Because of the wide range of execution time between queries, the figures in Section~\ref{sec:eval:opt} report ``speedup" (optimized execution time over baseline execution time) to normalize the results, but the full execution times for the baseline interpreted queries are reported in Table~\ref{tab:interp}.
\revA{We report both the speedup of the online-optimized program without indexes compared to the baseline without indexes, and the speedup of the online-optimized program with indexes compared to the baseline with indexes. Because indexing can reduce costly relation scans, the overall execution time is shorter when using indexes so the relative speedup is generally lower than the speedup without indexes.}
\input{06-eval/a-table}

\subsubsection{JIT-optimizing ``unoptimized" programs}\label{sec:eval:worst}
We first consider how close \system{} can get the performance of programs that are not already optimized to the performance of hand-optimized programs. 

In Figures~\ref{fig:maxi-worst} and ~\ref{fig:mini-worst}, the speedup of the JIT-optimized \system{} program over the ``unoptimized", interpreted input program is shown. 
\revA{The CSPA benchmark run on the full httpd dataset, when run without join order optimizations, can produce intermediate relations larger than the maximum value storable in a 32-bit integer. Because \system{} currently uses 32-bit integers to index into arrays, there exists a current limitation that \system{} cannot support relations with larger than the largest 32-bit integer. Custom array implementations or clustered indexes remain as future work, but to illustrate the example while avoiding this limitation, in Figures~\ref{fig:maxi-best} and ~\ref{fig:maxi-worst} we show the speedup of the CSPA query run on a sample of 20\,000 input tuples from the httpd dataset, indicated as CSPA\_20k. The execution time of the full httpd dataset is reported in Section~\ref{sec:eval:sota}. The CSDA and CSPA\_20k experiments were only run on \system{} with indexes due to the large runtime, and because the CSDA program has only 2-way joins we did not include an ``unoptimized" comparison as the runtime is identical to the ``hand-optimized".

The ``hand-optimized" CSPA\_20k program runs about ${\sim}$2000x faster than the ``unoptimized" program. In the best case, \system{}'s JIT optimizer produces a program with a speedup of around 
% 2886x %46
${\sim}$5000x %44
over the ``unoptimized" input program, with no input from the user. The IRGenerator backend also can outperform the ``hand-optimized" programs but less consistently than the Lambda backend. However for the CSDA program, because there are only 2-way joins the IRGenerator produces a much more significant speedup compared any of the backends that have more significant overheads.} The reason for this improved performance is that for the JIT the optimization is applied repeatedly, while for the hand-optimized program the same join order is used throughout the execution.

\noindent\textbf{Compilation Targets.} We compare the different backends to show the tradeoff between expressibility, maintainability, and performance. The Quotes backend is the most expressive and easy to maintain as the code generated is fully typed at compile-time by the Scala compiler. It also allows for finer-tuned application of optimizations and arbitrary runtime code generation. However, the overhead costs of invoking the compiler and waiting on the result can limit the speedup, for example for the Inverse Functions query with synchronous compilation, the Quotes speedup is only 
% 817x. %46
300x, compared to the ``potential" speedup of ${\sim}$2100x. \revA{This drops even further, to only 30x speedup, when indexes are used because the overhead of code generation stays constant while the overall time is much less for a program run with indexes.} %44
At the other end of the spectrum, the Bytecode backend more closely resembles traditional LLVM-IR generating RDBMSs and requires unsound generation of bytecode. The bytecode compilation target exchanges maintainability and correctness checks for performance, 
% as it performs better at 1267x but only allows for a single code generation stage. % 46
as it performs comparably to the Quotes backend on some programs, like CSDA, but performs worse on other programs, like the Ackermann function. %44
The Lambda backend relies on the JVM's ability to aggressively inline functions that are called repeatedly. While the Quotes and Bytecode backends will improve with more iterations, as the JVM warms up, they do not show the same sharp increase in performance as the Lambda backend does with repeated calls. However the Lambda backend can only call the set of predefined lambdas, so it cannot arbitrarily generate code at runtime the same way that the Quote and Bytecode backends can. The Lambda backend can even outperform the interpreted ``hand-optimized" program
 on some experiments %46
%, for example it performs 2424x better for the inverse function analysis. This is %44
due, in part, to the lack of tree traversal overhead. Lastly, the IRGenerator will regenerate the internal IR on the fly, which can then be passed to the interpreter. 
%For the specific case of the optimization described in Section~\ref{sec:opt}, it can be at least partially applied at the IR level; however, this approach would not scale for lower-level optimizations. 
As only the IROps are generated, the ``compilation" phase is minimal compared to the other backends, so the overhead of applying the optimization is minimal.

\noindent\textbf{Asynchronous Compilation.} 
Because \system{}'s ``safe points" occur between IROps, for nonblocking compilation, the overheads of invoking the compiler can delay the application of the optimization so the \system{} program may run un-optimized for a few iterations before the compiled code is ready. 
%In Figure~\ref{fig:all-tasty-worst}, for the bars marked ``async" \system{} does not wait for compilation to finish and instead continues interpreting the unoptimized program until the compiled, optimized program is ready. For the bars marked ``blocking", \system{} blocks evaluation until the optimized code is ready, so no unoptimized Datalog is ever run, but the overheads of invoking the compiler are unavoidable. 
For longer-running compilations, like when generating Quotes, blocking to wait for compilation can impact the overall speedup significantly. For example, for the inverse function benchmark, \system{} goes from a 
% 1354x %46
${\sim}$900x %44
speedup when asynchronously compiling to a 
% 817x %46
${\sim}$300x %44
speedup when blocking because more time is spent generating quotes than saved by the optimization.
% \system{} does not bother with asynchronous compilation for the Lambda or 
% For short-running compilations, like that of the Lambda backend, blocking compilation shows a 
% % 2519x %46
% 2424x %44
% speedup compared to the async speedup of only 
% % 1694x, %46
% 1700x, %44
% as no iteration is allowed to proceed without the optimization applied and the JVM has more opportunities to optimize the generated lambdas. 
For \revA{microbenchmarks}, 
% shorter-running programs that operate over smaller amounts of data, 
\system{}'s JIT optimizer can surpass the hand-optimized potential for asynchronous and blocking compilation as seen in Figures~\ref{fig:mini-worst}, but in some cases the asynchronous compilations may never be used if the execution of the interpreted subtrees are fast enough to finish before compilation is ready, and no speedup is seen. \revA{The use of indexes can even out the difference between the different backends, as the system is not bottlenecked on relation scans and so applying the optimizations has less impact.}

\subsubsection{JIT-optimizing ``hand-optimized" programs}\label{sec:eval:best}
We now consider how much \system{} can unintentionally \emph{degrade} the performance of programs that are already hand-optimized. 
In Figures~\ref{fig:maxi-best} and ~\ref{fig:mini-best}, we show the speedup (or slowdown) of applying \system{}'s JIT optimizations to already hand-optimized input Datalog programs. At worst, for very short-running \revA{microbenchmarks} where the cost of applying any optimization is a large proportion of the total runtime, \system{} shows an end-to-end slowdown of 
%0.2x when async and 0.01x when blocking %46
${\sim}$0.1x % 44
compared to the ``hand-optimized" program run on the interpreter for the Ackermann function.

\noindent\textbf{Compilation Targets.} In contrast to the results presented in Section~\ref{sec:eval:worst}, for already-optimized programs, longer-running asynchronous compilation can outperform shorter-running asynchronous compilation because it enables \system{} to more quickly apply an optimization that in fact produces a slightly slower program than the input program. For example, for Andersen's Points-To analysis on SListLib, the Lambda backend has a runtime of 
${\sim}$0.45x %both 44, 46
the ``hand-optimized" runtime. The Quotes backend, which has a higher compilation cost, has a slightly better runtime of 
% 0.4x %46
${\sim}$0.6x %44
the ``hand-optimized" program. 

How close the optimization can get to the ``hand-optimized" program, or if it can be exceeded, is a property of the program itself.
\revA{For example, the CSDA benchmark does not experience any performance degradation when \system{} applies JIT optimization to already-optimized programs using the Lambda backend. In fact, the IRGenerator program is ${\sim}$6x faster than the interpreted hand-optimized program because the overhead cost is low and optimizing 2-way joins only is the same as swapping the build and probe side of the join at runtime, as there is no selectivity estimation to affect the join order.}
%because the difference between hand-optimized and JIT-optimized is so small that the benefits of compiling and avoiding tree traversal outweigh the speedup of hand-tuning the input program. 

Ultimately, hand-tuning programs have consistently shown better performance than compiler-optimized programs but require intense effort and deep knowledge of Datalog and \system{}-specific implementation details, and are challenging to scale to large input programs. Yet on both CSPA\_20k and CSDA, online optimizations outperform the hand-tuned programs.

\subsection{Ahead-of-time Compilation}\label{sec:eval:aot}
% \vspace{-0.3}
%\system{}'s JIT applies the online join ordering optimization multiple times once query execution has begun.
\system{} can compile query plans ahead-of-time, when \system{} itself is compiled, reordering the join operations using the set of query facts and rules that are available. This \system{}-compile-time, or ``offline" optimization since the cost of compilation is not included in the query execution time, is implemented with macros that utilize the same backend as \system{}'s JIT Quote-compiler\cite{unifmsp}.
%(but at the compile time of \system{} as opposed to the query comile-time in the JIT). 
This offline sort optimization can be combined with continuous online join reordering by ahead-of-time compiling the IRGenerator backend. Because the execution engine uses Timsort\cite{timsort} to reorder relations, sorting already-sorted inputs is done in linear time, and sorting inputs that have sorted subsets will perform better than completely unsorted inputs. Thus, even if not exactly correct but close, presorting offline will contribute to quicker online sorting and is key to why compile-time sorting, even if facts are not fully available, is worth doing even if online sorting is enabled.
%and sorting near-sorted inputs will perform better than unsorted inputs.

In Figure~\ref{fig:aot}, we compare the speedup of query execution time over the ``unoptimized" interpreted query (Table ~\ref{tab:interp}) 
in the following configurations: ``JIT-lambda", \system{}'s JIT with the lambda backend at the lowest granularity, which does not have access to any facts or rules before execution begins; ``Macro Facts+rules (online)" which has access to all facts and rules at \system{} compile-time so will apply the sorting optimization using the input relational cardinalities. The generated code includes the online optimization using the IRGenerator. ``Macro Rules (online)" works similarly, except it only has access to rules and no fact relation cardinalities at compile-time. As a result, the reordering of relations will depend only on the selectivity of the rules and not on the relation cardinalities. This configuration also injects the online sort into the code generated at compile-time. ``Macro Facts+rules" does the compile-time reordering using all relations, but does not inject the online optimization. Finally, ``Macro Rules" does the compile-time reordering using only the rules and does not generate code with the online optimization applied. In Figure~\ref{fig:aot}, experiments that include continuous relation reordering after query execution begins are shown with a \revA{dotted}-line border. Orthogonally, experiments that access no fact or rule information at \system{} compile time are shown with a solid color, those that access rules only are spotted, and those that access rules and facts are striped.
%TODO: need to specify why we chose short-running queries here? A: shorter running queries are more sensitive to the configuration changes compared to the longer-running one. 

% online vs offline
Generally, online reordering optimizations perform better than offline optimizations because the order of relations fed to the join can continuously update. However, continuously reordering relations has some overhead, and for programs for which the relation cardinalities do not change dramatically in comparison to each other, a single join order over the course of query execution may perform better than online reordering due to avoiding that overhead (illustrated in the case of the Primes benchmark). 

% lambda vs. macro
The JIT-lambda configuration performs comparatively, or sometimes better than the macros+online IRGenerator optimization because the lambda-generated code avoids the tree-traversal overhead present in the macro configurations, as the macros are essentially compiling the interpreter.

% facts vs. facts+rules 
The macro experiments that gain access to fact relation cardinality at compile-time tend to perform better than those that rely only on rule selectivity, due to the cost of ingesting facts at runtime and a more optimal join-order due to knowledge of the relation cardinalities. However this is not always the case, for example in the Fibonacci benchmark, the Rules-only macro slightly outperforms the Macro Facts+Rules, a result of the join-ordering algorithm privileging relation cardinality over selectivity where, in reality, the selectivity is more influential on the overall execution time. 

% In Figure~\ref{fig:macro-large} and ~\ref{fig:macro-small} the speedup of the interpreted hand-optimized programs is compared with the speedup of compile-time-only optimizations (``Macro Rules" and ``Macro Facts+rules", in the same configuration as the corresponding bars in Figure~\ref{fig:aot}). While not as effective as runtime-optimized programs, compile-time optimized programs show at best a 1100x speedup for the Inverse Functions query, and can even slightly out-perform the hand-optimized programs for \revA{microbenchmarks} like Equal, Fibonacci, and Primes, in part due to the use of macros.

To conclude, for most but not all benchmarks, the earlier information is known, the better the optimizations applied, and combining compile-time and run-time optimizations perform better than each individually. Whether or not these conclusions hold for specific benchmarks is a property of the query itself and how relation cardinalities change over the course of query execution. These properties are difficult to predict statically, especially without access to the full fact and rule relations, so queries with varying properties are presented in this subsection. However, regardless of the query, all presented configurations outperform the baseline, unoptimized queries.

\input{06-eval/macro-figs}

\subsection{Comparison with State-of-the-art}\label{sec:eval:sota}

% \begin{figure}
% \centering
% % \frame{
%   \includegraphics[trim={1.9cm 18.5cm 5.6cm 2.2cm}, width=\linewidth,clip]{../figures/sota.pdf}
%   % \vspace{-1.6cm}
%   \caption{Execution time Soufflé and Carac}\label{fig:sota}
%   % \vspace{-0.8cm}
% \end{figure}

%The goal of \system{} is not to operate as a Datalog execution backend, nor as a set of predefined static analyses, but as a framework to compose interactive metaprogramming queries embedded within a general-purpose programming language. These queries can be called within and leverage the internal optimizations of analytical pipeline systems. However, 
To get a sense for how \system{} compares against existing Datalog engines, in this section we compare the execution time of equivalent \system{} queries with those of a state-of-the-art Datalog execution engine.

One current state-of-the-art for Datalog evaluation is Soufflé~\cite{souffle1}, which ingests an extension of Datalog and transforms the input program through a series of partial evaluations, generating template-heavy C++ programs that get ultimately compiled into a binary executable. Soufflé and \system{} both generate an imperative program specialized to the input Datalog program in the form of an intermediate representation comprising of relational algebra operators, control flow statements, and relation management operations. Soufflé supports interpretation mode and compilation mode, while \system{}'s JIT  % visits the generated internal program representation and 
alternates, depending on runtime information, between interpretation of the intermediate representation and code generation using Multi-Stage programming. Further, \system{} supports a Datalog-based DSL embedded in the Scala language that allows for interoperability with Scala itself. In contrast, the Soufflé language is not an embedded DSL but its own Datalog-based language with extensions, including aggregation, user-defined functors, macros, and others.

Soufflé provides auto-tuning capabilities via a profiling phase\cite{souffle2} that identifies optimal join orders. %by recording relation cardinalities. 
%Soufflé implements extensive Datalog optimizations, including a specialized B-tree implementation, auto-index selection, Magic-Set transformations, and others. 
%Additionally, the language supported by Soufflé is more complex than that of \system{}, requiring a more complex type system. Lastly, Soufflé is a programming language, while \system{} supports a DSL tightly embedded within Scala. 
%DIASCLD46

\begin{table}[htbp]
\caption{Average Execution time (s) of DLX, Soufflé, and Carac}
\centering
%\resizebox{0.95\linewidth}{!}{
%\resizebox{\columnwidth}{!}{%
\begin{tabular*}{\columnwidth}{
    l|
    S[table-format=2.1]|
    S[table-format=3.2]|
    S[table-format=3.0]|
    S[table-format=3.0]|
    S[table-format=4.2]
}
%\hline
 & {\parbox{0.7cm}{\centering DLX}} & {\parbox{1.1cm}{\centering Soufflé \\ Interpreter}} & {\parbox{1cm}{\centering Soufflé \\ Compiler}} & {\parbox{1.4cm}{\centering Soufflé\\ Auto-Tuned}} & {\parbox{1cm}{\centering Carac \\ JIT}} \\
%\multicolumn{5}{|c|c|c|c|c|c]{\multicolumn{1}{|c|}{} & \multicolumn{1}{c|}{DLX} & \multicolumn{3}{c|}{Souffle} & \multicolumn{1}{c|}{Carac JIT}} \\
%\multicolumn{6}{|c|c|c|c|c|c}{a& {DLX} & {Souffle} & {Souffle} & {Souffle} & {Carac JIT}} \\
%\cline{2-6}
%\multicolumn{1}{|c|}{} & \textbf{} & \textit{Interpreter} & \textit{Compiler} & \textit{Auto-Tuned Compiler} & \textbf{} \\
%\multicolumn{1}{|c|}{} & \multicolumn{1}{c|}{DLX} & \multicolumn{1}{|p{3cm}|}{Souffle \\ Interpreter} & \multicolumn{1}{|p{3cm}|}{Souffle} & \multicolumn{1}{|p{3cm}|}{Souffle} & \multicolumn{1}{c|}{Carac JIT} \\
% & {DLX} & {Souffle Interpreter} & {Souffle Compiler} & {Souffle Auto-Tuned Compiler} & {Carac JIT} \\
\hline
InvFuns & 2.60 & 0.16 & 23.39 & 26.48 & 0.04 \\
\hline
CSDA & 33.87 & 13.15 & 22.26 & 22.44 & 64.08 \\
\hline
CSPA & {DNF} & 247.65 & 195.43 & 271.85 & 4143.49 \\
\hline
\end{tabular*}
%}
\label{tab:sota}
\vspace{-1pt}
\end{table} 

\revA{We also compare to a commercial Datalog engine (DLX), anonymized
due to their license agreements.}
Table~\ref{tab:sota} shows the execution time of various programs expressed in Soufflé, \revA{DLX} and \system{}, on Soufflé version 2.4 and \system{} in ``full" mode, run synchronously at a granularity that takes into account delta relations ($\sigma\pi\Join$). All commercial engines are invoked without any special arguments beyond silencing warnings, and the location of profiling files for the experiments that include a profiling phase.

In Table~\ref{tab:sota}, "Soufflé Auto-Tuned" refers to Soufflé in compiler mode using a profile generated ahead-of-time on the same data we use for benchmarking.
\system{} outperforms Soufflé when compiling, both with and without the profiling phase included, for the Inverse Functions ("InvFuns") experiment because \system{} can avoid the cost of invoking a full C++ compilation, which dominates Soufflé's execution time. It outperforms the interpreter by ${\sim}$4x, and the compiler by ${\sim}$600x. \revA{\system{} outperforms DLX by ${\sim}$65x on the Inverse Functions experiment.} \system{} performs better relatively for Datalog queries with longer atoms and more complex join conditions.

For longer-running programs with larger input datasets, Soufflé's best run time outperforms \system{} best run time by ${\sim}$5x for CSDA and ${\sim}$20x for CSPA. \revA{DLX outperformed \system{} by ${\sim}$2x on CSDA, but did not finish the CSPA experiment within a 24-hour window.}

%For \revA{microbenchmarks} like Ackermann and CBA, \system{} (Lambda, warmed-up) outperforms Soufflé's interpreter without profiling (``preprofiled-interp") by 14x and 32x respectively, and with profiling (``profile-interpreter") by 30x and 64x respectively. For Points-To and Inverse Functions, Soufflé's ``preprofiled-interp" outperforms \system{}'s best execution time (Lambda, warmed-up) by 3.3x and 3.6x. With the profiling phase included, Soufflé's interpreter outperforms \system{} by 1.4x and 1.5x. 

However, this is an imperfect comparison because the Soufflé variant of Datalog is more expressive and may spend additional cycles for language features that the programs do not utilize, which would help \system{} outperform Soufflé for some experiments. Additionally, Soufflé implements many orthogonal optimizations that \system{} does not currently support, for example specialized B-tree implementation, relation inlining, and others that would help Soufflé outperform \system{} for other experiments. Although the comparison is not perfect, it can be concluded from Figure~\ref{tab:sota} that for the longest-running of the presented programs, Soufflé's best auto-tuned execution time outperforms \system{}'s best execution time by at most ${\sim}$20x, but does not include the time spent generating the profiling information in order to auto-tune.

Ultimately, the goal of \system{} is not to present an alternative to commercial Datalog engines but to illustrate a potential runtime optimization that leverages advanced programming language capabilities like quotes \& splices that would enable Soufflé to avoid a profiling phase entirely, which for large programs can be significant.
\system{} can tightly knit a popular general-purpose programming language, Scala, with declarative Datalog queries while still delivering execution times that are competitive with the most advanced Datalog systems.

%% file: 06-eval/figs.tex
\begin{figure}
\centering
  % \frame{
  \includegraphics[trim={1.9cm 20.5cm 5.8cm 1.9cm}, width=\linewidth,clip]{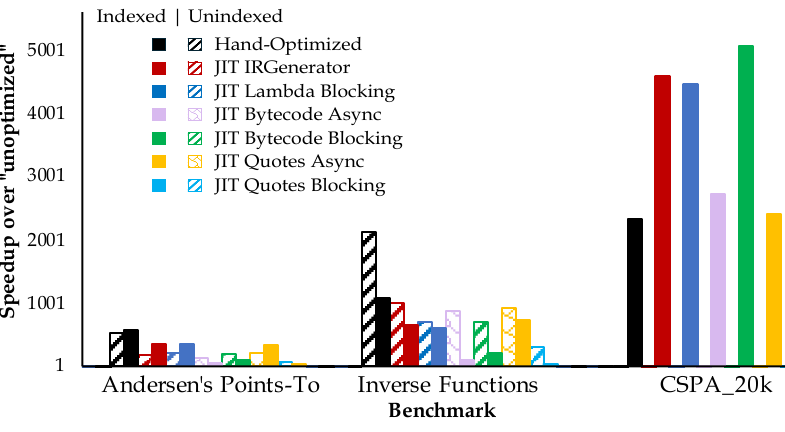}%}
  % }
  \vspace{-.2cm}
  \caption{Macrobenchmarks compared to unoptimized}\label{fig:maxi-worst}
    \vspace{-0.3cm}
\end{figure}

\begin{figure}
\centering
  % \frame{
  \includegraphics[trim={1.9cm 20.8cm 6cm 1.9cm}, width=\linewidth,clip]{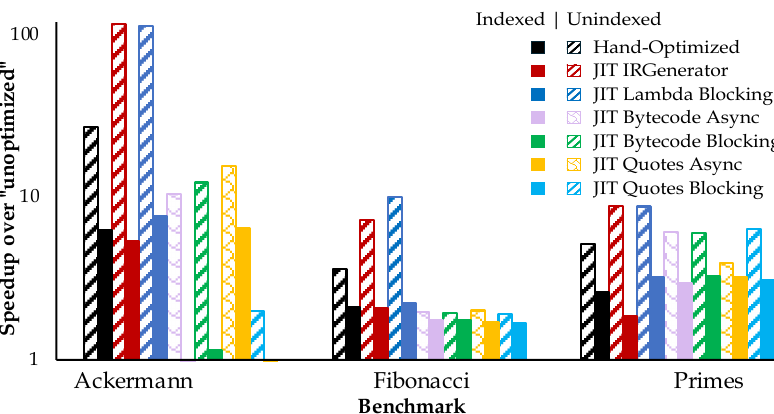}
    % }
  \vspace{-0.8cm}
  \caption{Microbenchmarks compared to unoptimized (log scale)}\label{fig:mini-worst}
  \vspace{-0.2cm}
\end{figure}

\begin{figure}
\centering
  % \frame{
  \includegraphics[trim={1.9cm 20.5cm 5.5cm 2.2cm}, width=\linewidth,clip]{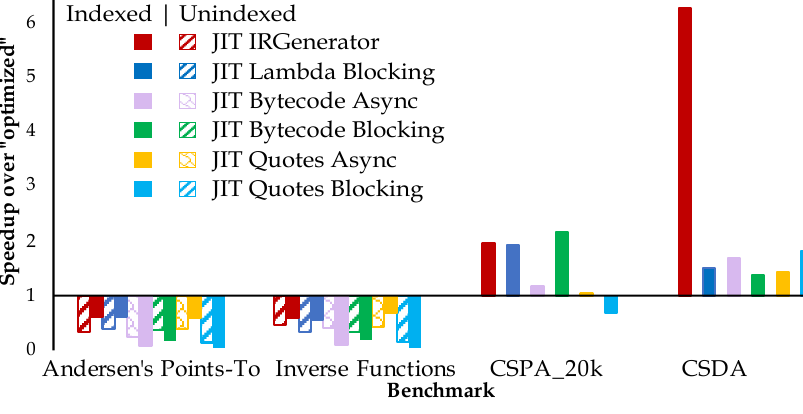}
    % }
  \vspace{-.8cm}
  \caption{Macrobenchmarks compared to hand-optimized}\label{fig:maxi-best}
  \vspace{-0.3cm}
\end{figure}

\begin{figure}
\centering
  % \frame{
  \includegraphics[trim={1.9cm 20.5cm 5.5cm 2.1cm}, width=\linewidth,clip]{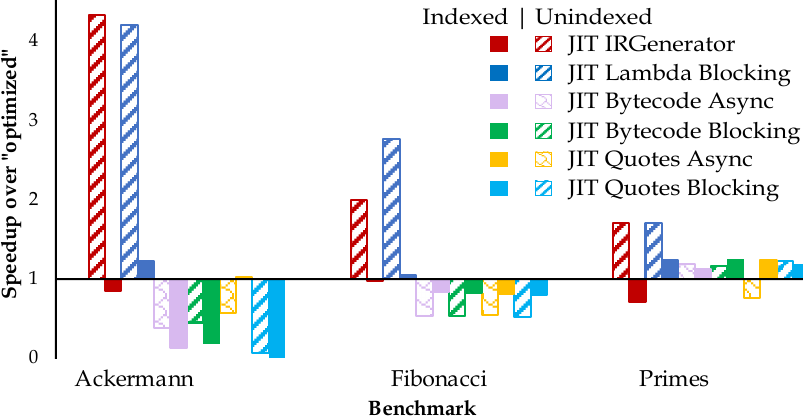}
    % }
  \vspace{-.8cm}
  \caption{Microbenchmarks compared to hand-optimized}\label{fig:mini-best}
  \vspace{-0.3cm}
\end{figure}

%% file: 06-eval/a-table.tex
% \begin{table}[htbp]
% \caption{Execution time (s) of interpreted Carac queries}
% \centering
% \setlength{\tabcolsep}{12pt} % Adjust the column spacing here to control the width of the table
% \begin{tabular}{|c|c|c|}
% \hline
% \textbf{Baseline Query} & \textbf{Optimized} & \textbf{Unoptimized} \\
% \hline
% Ackermann & 0.272208 & 7.001168 \\
% \hline
% CBA & 0.008064 & 0.038313 \\
% \hline
% Equal & 0.019881 & 0.092675 \\
% \hline
% Fibonacci & 0.102191 & 0.336177 \\
% \hline
% Primes & 0.015402 & 0.103381 \\
% \hline
% Points-To & 12.246468 & 7229.473415 \\
% \hline
% Inverse Functions & 7.504203 & 16980.17118 \\
% \hline
% \end{tabular}
% \vspace{3pt}
% \label{tab:interp}
% \end{table}

\begin{table}[htbp]
\caption{Average Execution time (s) of interpreted Carac queries}
\centering
%\resizebox{0.95\linewidth}{!}{
\begin{tabular*}{\columnwidth}{
    l|
    S[table-format=3.3]|
    S[table-format=1.4]|
    S[table-format=5.4]|
    S[table-format=3.4]
}
%\hline
%\multicolumn{1}{|c|}{} & \multicolumn{2}{c|}{Unindexed} & \multicolumn{2}{c|}{Indexed} \\
%\cline{2-5}
 & \parbox{1.4cm}{\centering Unindexed\\ Unoptimized} & \parbox{1.4cm}{\centering Unindexed\\ Optimized} & \parbox{1.4cm}{\centering Indexed\\ Unoptimized} & \parbox{1.2cm}{\centering Indexed\\ Optimized} \\
% & {\parbox{0.7cm}{\centering DLX}} & {\parbox{1.1cm}{\centering Soufflé \\ Interpreter}} & {\parbox{1cm}{\centering Soufflé \\ Compiler}} & {\parbox{1.4cm}{\centering Soufflé\\ Auto-Tuned}} & {\parbox{1cm}{\centering Carac \\ JIT}} \\
\hline
Ackermann &     0.2164 &     0.0080 &     0.0110 &     0.0017 \\
\hline
Fibonacci &     0.0116 &     0.0032 &     0.0018 &     0.0009  \\
\hline
Primes &        0.0029 &     0.0006 &    0.0011 &     0.0004 \\
\hline
Andersen &      159.6990 &   0.3005 &     14.8536 &    0.0258 \\
\hline
InvFuns &       693.8857 &   0.3264 &     31.3428 &    0.0289 \\
\hline
CSDA & {--} & {--} &                    435.2269 &   435.2269 \\
\hline
CSPA\_20k & {--} & {--} &               19777.1323 & 8.4743 \\
\hline
\end{tabular*}
%}
\label{tab:interp}
\vspace{-1pt}
\end{table}

%% file: 06-eval/macro-figs.tex
\begin{figure}
\centering
% \frame{
  \includegraphics[trim={1.9cm 22cm 5.8cm 1.9cm}, width=\linewidth,clip]{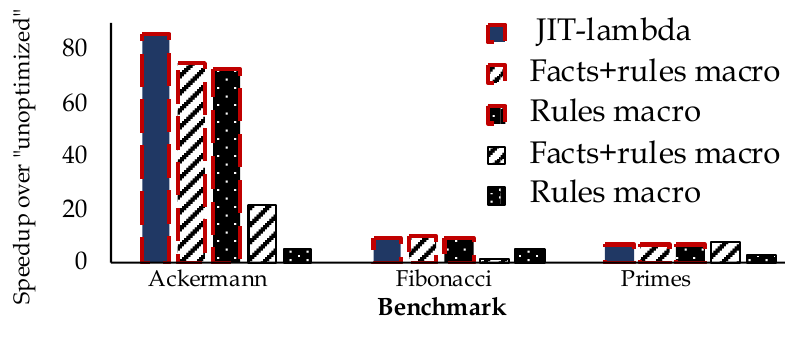}
  \vspace{-0.8cm}
  \caption{Microbenchmarks: Ahead-of-time and Online Compilation}\label{fig:aot}
  \vspace{-0.6cm}
\end{figure}

% \begin{figure}
% \centering
% % \frame{
%   \includegraphics[trim={1.9cm 20.5cm 5.8cm 1.9cm}, width=\linewidth,clip]{../figures/macro-offline-small.pdf}
%   % \vspace{-1.6cm}
%   \caption{Microbenchmarks: Ahead-of-time Only Compilation }\label{fig:macro-small}
%   % \vspace{-0.8cm}
% \end{figure}

% \begin{figure}
% \centering
% % \frame{
%   \includegraphics[trim={1.9cm 20.5cm 5.8cm 1.9cm}, width=\linewidth,clip]{../figures/macro-offline-tasty.pdf}
%   % \vspace{-1.6cm}
%   \caption{Macrobenchmarks: Ahead-of-time Only Compilation}\label{fig:macro-large}
%   % \vspace{-0.8cm}
% \end{figure}

%% file: 07-related-work/main-v0.tex
% \vspace{-0.1cm}
Program optimization and runtime analysis have received attention from both the database and compiler communities. % This section surveys the related work. % and explains how \system{} extends the state-of-the-art.

\noindent \textbf{Datalog}. Datalog has been used to solve fixed-point problems within programming languages, for example as the basis of Ascent, a Datalog-based logic programming language embedded in Rust for use in the borrow checker~\cite{rustdl}.  Datalog has also recently been used for Java pointer and taint analysis in Doop~\cite{doop}, TensorFlow~\cite{tensorflow}, Ethereum Virtual Machine~\cite{etherium}, and AWS network analysis~\cite{aws}.  Datalog is the logical foundation for Dedalus~\cite{dedalus}, a language designed to model distributed systems upon which the Bloom language is founded.  Bloom uses Datalog as a basis to formally guarantee consistency properties of distributed systems.  The HYDRO language stack~\cite{ndicp} takes this approach a step further and argues that the next programming paradigm for cloud computing should involve using verified lifting to translate user-written programs into a logic-based internal representation language that has roots in Datalog.  Flix~\cite{flix16} is a declarative programming language that integrates first-class Datalog constraints into the language. Several Datalog engines do not implement any join-reordering optimizations\cite{datalogmodern}, but RecStep\cite{recstep} optimizes Datalog subquery execution at every iteration by using runtime statistics, including reading relation cardinalities and switching the build and probe sides for their hash-join implementation. \system{}'s approach differs in that it leverages code generation and reorders the entire \emph{n}-way join once per-iteration, per-predicate, or per-subquery, not per 2-way join.

% \paragraph{Queries over Code} SQL or Declarative SQL-like queries have been used to query over program traces in the Program Trace Query Language (PTQL)~\cite{rqopt} to instrument code while profiling and debugging.  Stack graphs\cite{stackgraphs} enable name resolution across programming languages for precise navigation within GitHub by encoding name binding information about a program in a graph structure that can be queried using pathfinding algorithms.  Object managers like Microsoft Repository ~\cite{bernstein1999microsoft} store object-oriented program metadata in Microsoft SQL Server, providing SQL query support.  Databases have been used to implement low-code systems like Oracle HTML DB~\cite{htmldb} and Oracle Apex ~\cite{oracleapex} that enable developers to assemble an HTML interface (or application) on top of the database, which stores the programs in tables.  External libraries built on top of compilers, like Clang Tidy or Clang Static Analyzer\cite{clanganalyzer}, enable a set of predefined static analyses to be performed on information generated by the compiler.  CodeQL~\cite{codeQL} and SemGrep~\cite{semgrep} are analysis engines that operate over code to automate security checks and perform variant analysis.  These tools provide powerful mechanisms for answering specific types of queries (for example name resolution, security analysis, etc.). %but are not easily composed to declaratively construct arbitrary queries for new analyses.

\noindent \textbf{Adaptive Query Processing}.  % Thus, there is a missed opportunity for optimizations that may only become possible when runtime information becomes available and entirely avoids the complexities of cardinality estimation and heuristic-based cost models for query optimization. 
Just-in-time code-generating database systems like HyPer, HIQUE, Hekaton, ViDa, and many others~\cite{vectorized,adaptiveexec,vida} use SQL as a declarative specification to generate programs at query compile-time.  There are several JVM-based bytecode-generating DBMSs, for example Presto\cite{presto} and Spark\cite{spark}.  Generally, once the code for the queries is generated, they are not re-generated within the execution of a single query by the generated program.  Adaptive Query Processing (AQP), however, which uses runtime statistics to inform query evaluation decisions to avoid the well-known inaccuracy of cost model-based estimation~\cite{costmodels}, has received significant attention from the database community~\cite{aqp}.  Notably, NoisePage~\cite{pcq} and Umbra~\cite{dblocks} combine AQP with code generation to modify generated query plans at runtime: NoisePage with manipulation of function pointer arrays and Umbra by embedding code fragments for all possible execution variants into the generated code and using LLVM's \texttt{indirectbr} instruction to jump to label addresses at runtime.  Most code-generating AQP systems target general-purpose SQL, while the recursive nature of Datalog queries makes the primary considerations for query optimization different~\cite{souffle2}, so Soufflé uses runtime information to inform join optimization decisions but requires an offline profiling stage.
% Most Datalog execution engines are limited to compile-time code generation that does not use runtime information from the current execution to further specialize code once it has been generated.% \system{} code generation is done with Multi-Stage metaprogramming language constructs, so it is not limited to preexisting function definitions or reliant on instruction addresses, is fully portable, and is not limited to a single code generation phase: in fact, one can continue to generate code in new stages an arbitrary number of times. 
\system{}'s goal of applying optimizations at the level of abstraction most appropriate for the optimization is shared by MLIR~\cite{mlir}, an intermediate representation that allows co-optimizing multiple representation levels. Similarly to MLIR and verified lifting, \system{} follows a multi-representation strategy and lifts input programs but additionally incorporates runtime information to generate optimal execution strategies. 

%% file: 08-conclusion/main-v0.tex
% \vspace{-0.1cm}
% \TODO{rework}
% Trends in modern software development foreshadow an explosive growth in both the volume and complexity of software, far beyond the capacity of humans to manage, optimize, and debug. It may even seem that the ``end of programming" is nigh, and soon engineers will train, not write, their programs~\cite{endofprog}. Yet until this generated code (however convincing) can be proven correct, the utility of these tools in developing mission-critical applications is limited and may lead to fatal bugs or require time-consuming manual optimization by domain experts.
%Testing can identify errors in runtime behavior but cannot guarantee that no errors exist~\cite{dijkstra}. On the other end of the spectrum, formal methods may provide stronger guarantees than necessary and may not scale efficiently. 
% In this work, we make the case for \arch{}, a novel %``database for code"
% methodology for intuitive, online analytical queries over programs at scale. We show how to apply \arch{} in \system{} by tightly embedding a specialized Datalog execution engine with %the host programming language, 
% Scala. \system{} leverages novel language features to apply continuous optimizations that speed up both unoptimized and hand-optimized programs.
In this work, we introduced \arch{}, a novel approach that combines the power of just-in-time code-generating data management systems with principled metaprogramming to optimize recursive queries at compile and runtime. Unlike existing methods, \arch{} enables code generation and optimization to continue past the start of query execution, potentially deoptimizing when necessary, to better adapt to real-time data changes. 
% Through \system{}, we show how \arch{} can speed up unoptimized Datalog queries by three orders of magnitude by optimizing the join order of conjunctive queries within Datalog execution and a 4x speedup over hand-tuned queries while maintaining competitive execution times with the state-of-the-art Datalog systems.
Through \system{}, we demonstrate that \arch{} can enhance the performance of Datalog queries substantially, achieving a speed increase of three orders of magnitude over unoptimized queries and 6x over hand-tuned queries by optimizing the join order of conjunctive queries during Datalog execution. 

%\system{} can tightly knit a popular general-purpose programming language, Scala, with declarative Datalog queries while still delivering execution times that are competitive with the most advanced Datalog systems.